\documentclass[amssymb,amsmath,aps,pre,preprint,showpacs]{revtex4} 
\usepackage{graphicx}

\def\func#1{\textrm{#1}}

\begin{document}

\title{Atomic-scale structure of hard-core fluids under shear flow}
\author{James F. \surname{Lutsko}}
\affiliation{Center for Nonlinear Phenomena and Complex Systems\\
Universit\'{e} Libre de Bruxelles\\
Campus Plaine, CP 231, 1050 Bruxelles, Belgium}
\email{jlutsko@ulb.ac.be}
\date{\today }

\begin{abstract}
The effect of velocity correlations on the equal-time density autocorrelation function, 
e.g. the pair distribution function or pdf, of a hard-sphere fluid undergoing shear flow is investigated.
The pdf at contact is calculated within the Enskog approximation and is shown to be in good agreement with
molecular dynamics simulations for shear rates below the shear-induced ordering transition. These calculations
are used to construct a nonequilibrium generalised mean spherical approximation for the pdf at finite separations
which is also found to agree well with the simulation data.
\end{abstract}

\maketitle

\affiliation{Center for Nonlinear Phenomena and Complex Systems.\\
Universite Libre de Bruxelles\\
B1050 Bruxelles\\
Belgium}

\section{Introduction}

In equilibrium, simple fluids exhibit spatial correlations which are
characterized by the pair distribution function (pdf) describing the
probability of finding two atoms with a give relative orientation and
separation. Equilibrium liquid state theory is primarily concerned with the
calculation of the pdf and a number of successful approaches have been
developed including the Percus-Yevik approximation for hard-spheres, the
mean-spherical approximation and the more recent self-consistent integral
equations\cite{HansenMcdonald}. Knowledge of the pdf is equivalent to
knowledge of the density-density static correlation function\cite%
{HansenMcdonald} and once this is known, all other interesting static
correlation functions, e.g. density-energy, energy-energy, ..., are
immediately known because the velocity-dependence of the two-body
distribution function in equilibrium is trivial. It is a characteristic of
nonequilibrium fluids that this property no longer holds\cite{Lutsko96,Lutsko2001,LutskoHCS,SotoMarechal},
 and the presence of
velocity correlations is the reason that the determination of static
correlations in nonequilibrium fluids, over all densities and length scales,
is a difficult problem.

The velocity correlations that occur in nonequilibrium fluids, as well as in
fluctuations about the equilibrium state, are generated by collisions which
have the effect of altering the two-body probability distribution so that
even if the velocities of the atoms prior to a collision are assumed to be
independent variables, the velocities after a collision are not independent.
The question of whether the of velocities of two atoms prior to a collision
are really independent variables has been much studied in statistical
mechanics over the last 30 years and phenomena such as long-time tails and
long-ranged correlations are proof that this assumption is not strictly
adhered to\cite{DorfKirkSeng,KirkBelSeng}, although in many cases it remains a good approximation. While
the calculation of the pre-collisional correlations is a very difficult
problem, it has recently been noted\cite{Lutsko96} that, for fluids
interacting via a hard-core potential, it is possible to calculate the
post-collisional correlations in an arbitrary nonequilibrium state for the
special case of the two atoms being in contact using the same assumptions as
underlie the Enskog theory of the one-body distribution function. This
allows one to calculate all static correlation functions for two atoms in
contact up to this level of approximation. It was subsequently shown that
this information could be combined with a formalism borrowed from
equilibrium liquid-state theory to create a successful model of the pair
distribution function of a granular fluid\cite{LutskoHCS} (i.e., a fluid of
inelastic hard-spheres). The purpose of the present paper is to describe an
extension of this model to inhomogeneous systems and to examine its
application to the particular case of a fluid of elastic hard-spheres
undergoing uniform shear flow\ (USF) and to present detailed comparisons of
the theory to the result of molecular dynamics simulations. Uniform shear
flow, in which the velocity in one Cartesian direction varies linearly with
position along another axis, is a particularly interesting example since the
density-density correlation function can be studied experimentally by means
of light scattering \cite{Ackerson}. Furthermore, the hard-sphere model is 
generally accepted as a reasonable analogy to certain types of colloidal suspensions, see for example ref. \cite{Pusey}
and references therein,  
for which it is possible to achieve conditions of strong shear (e.g., shear rates
comparable to the mean free time of the colloidal particles) in the laboratory which are otherwise 
inaccessible in simple fluids. 

The second section of this paper reviews the theory behind the calculation
of static correlations at contact and evaluates the density-density
correlation function at contact for the special case of USF. This makes use
of recent work on solution of the Enskog equation for high shear rates\cite%
{Lutsko_EnskogPRL,LutskoEnskog} to extend an earlier calculation\cite{Lutsko96} resulting in
an explicit expression for the pdf at contact in a sheared fluid. The third
section deals with models for the pdf at finite separations. It reviews two
well known theories which are potentially applicable to atomic
length-scales: the kinetic model studied by Hess and co-workers\cite{Hess85}
and the Langevin model of Ronis\cite{RonisShear}. The former involves an
undetermined parameter which, if the theory is to apply to atomic
length scales, can now be fixed by requiring agreement with the calculations
for two atoms in contact. The latter, while not involving any \textit{a
priori} unknown parameters is nevertheless phenomenological and a diffusion
constant appearing in its formulation has in fact been treated as a free
parameter when comparing to experiment\cite{Ackerson}. Again, it is noted
that the parameter can be fixed unambiguously by requiring agreement with
the calculated value at contact. It is also shown that these two theories are
in fact very closely related not withstanding their different motivations.
Finally, in this section the nonequilibrium version of the Generalized Mean
Spherical Approximation (GMSA) is introduced as a means of modeling the pdf at finite
separations based on the atomic-length scale information coming from the calculations of Section
2 and, qualitatively, the large-separation (i.e., small wave-vector) information provided by
mode-coupling theories, based on either kinetic theory\cite%
{TK_LightScattering,Hess85,Mirim} or fluctuating hydrodynamic%
\cite{Tremblay,Garcia-Colin,RonisShear,Lutsko_Fluctuations}, of which the Ronis theory is an example. 
This is not unlike the original motivation of Weisman in introducing the
equilibrium GMSA as a means of improving on the Percus-Yevik approximation by
incorporating accurate knowledge about the pdf at contact, from the Carnahan-Starling equation of state
and the pressure equation, to construct a  model of
the pdf for an equilibrium hard-core fluid accurate over a wide range of densities%
\cite{WaismanGMSA}. Recent work by Yustes and Santos\cite{Yuste91,Yuste93,Yuste94}, as well
as Carraro and Ciccariello\cite{CC}, has shown first that the Percus-Yevik approximation may
be viewed as, in some sense, the simplest approximation that provides
certain analytic properties that any distribution function must satisfy and
second, that the GMSA of Weisman may be viewed as a framework for
systematically extending this model so as to incorporate additional
constraints. It is with this motivation that the extension of the GMSA to
nonequilibrium systems was proposed\cite{Lutsko2001} as a means of modeling
the density-density correlation function, i.e. the pdf, at all length scales.

In section 4, these calculations are compared with the results of molecular
dynamics simulations over a wide range of shear rates and densities. As noted
previously\cite{Lutsko96}, there seems to be a strong correlation between
the rapid decrease, with increasing shear rate, of the pdf in certain
directions and the onset of shear-induced ordering of the fluid. Below this
transition, it is shown that the Enskog calculations of the pdf at contact
are quite accurate at small shear rates and low densities and becomes
increasingly inaccurate as the density and/or the shear rate increases. The
theories for the pdf at finite separations are also compared to MD and it is
found that all three theories are in qualitative agreement with the GMSA
providing the best quantitative agreement with simulation. The paper ends
with a discussion of the prospects to extend these results to other systems.
A preliminary description of some of these results has appeared previously%
\cite{Lutsko2001}.

\section{Theory of correlations at contact}

\subsection{Hard-sphere statistical mechanics}

Consider a system of $N$ elastic hard spheres of
diameter $\sigma $ in a cubic volume $V=L^{3}$ described by a Cartesian
coordinate system with axes $\widehat{\mathbf{x}},\widehat{\mathbf{y}}$ and $%
\widehat{\mathbf{z}}$. The boundary conditions will be discussed below. The
dynamics of the atoms consist of free-streaming, subject to the boundary
conditions, interrupted by elastic collisions. Two atoms having coordinates $%
\mathbf{q}_{i},\mathbf{p}_{i}$ and $\mathbf{q}_{j},\mathbf{p}_{j}$ at time $%
t_{0}$ will collide at time $t_{-}$ provided that $\sigma =\left| \mathbf{q}%
_{ij}(t_{-})\right| $, where $\mathbf{q}_{ij}(t_{-})=\mathbf{q}_{i}(t_{-})-%
\mathbf{q}_{j}(t_{-})$ and provided $\mathbf{q}_{ij}(t_{-})\cdot \mathbf{p}%
_{ij}(t_{-})<0$. Immediately after the elastic collision, the momenta become%
\begin{eqnarray}
\mathbf{p}_{i}(t_{+}) &=&\mathbf{p}_{i}(t_{-})-\widehat{\mathbf{q}}%
_{ij}(t_{-})\left( \widehat{\mathbf{q}}_{ij}(t_{-})\cdot \mathbf{p}%
_{ij}(t_{-})\right)  \label{collide} \\
\mathbf{p}_{j}(t_{+}) &=&\mathbf{p}_{j}(t_{-})+\widehat{\mathbf{q}}%
_{ij}(t_{-})\left( \widehat{\mathbf{q}}_{ij}(t_{-})\cdot \mathbf{p}%
_{ij}(t_{-})\right)  \notag
\end{eqnarray}%
so that the relative momentum is reversed along the line of contact and the
total momentum is unaffected.

The statistical description of the system is characterized by the N-body
distribution, $\rho _{N}(x_{1},x_{2}...x_{N};t)$ which gives the probability
of finding the system at a given phase point, where atom $1$ has phase $%
x_{1}\equiv (\mathbf{q}_{1},\mathbf{p}_{1})$ etc., at the specified time $t$%
. Its evolution is specified by the pseudo-Liouville equation%
\begin{equation}
\left[ \frac{\partial }{\partial t}+\sum_{i=1}^{N}\mathbf{p}_{i}\cdot \frac{%
\partial }{\partial \mathbf{q}_{i}}+\sum_{i<j}\overline{T}%
_{-}(ij)+\sum_{i=1}^{N}\frac{\partial }{\partial \mathbf{p}_{i}}\cdot 
\mathbf{F}(x_{i})\right] \rho _{N}(x_{1},x_{2}...x_{N};t)=0
\label{Liouville}
\end{equation}%
where the collision operator is given by\cite{McLennan} 
\begin{equation}
\overline{T}_{-}(ij)=\int d\mathbf{q}_{ij}\;\delta \left( q_{ij}-\sigma
\right) \left[ \widehat{b}_{ij}-1\right] \Theta \left( -\mathbf{q}_{ij}\cdot 
\mathbf{p}_{ij}\right)
\end{equation}%
with the effect of the momentum transfer operator $\widehat{b}_{ij}$ on an
an arbitrary function is to replace the relative momentum $\mathbf{p}_{ij}$
by its post-collisional value $\mathbf{p}_{ij}-2\widehat{\mathbf{q}}%
_{ij}\left( \widehat{\mathbf{q}}_{ij}\cdot \mathbf{p}_{ij}\right) $ (see Eq. %
\ref{collide}). The final term of Eq. (\ref{Liouville}) describes any
external one-body forces acting on the atoms. Integrating (\ref{Liouville})
over $N-n$ of the coordinates yields the n-the equation of the BBGKY
hierarchy which the relates the $n$-body distribution to the $n+1$-body
distribution. In particular the result of choosing $n=N-1$ is 
\begin{eqnarray}
\left[ \frac{\partial }{\partial t}+\mathbf{p}_{1}\cdot \frac{\partial }{%
\partial \mathbf{q}_{1}}+\frac{\partial }{\partial \mathbf{p}_{1}}\cdot 
\mathbf{F}(x_{1})\right] \rho _{1}(x_{1};t) = \label{BBGKY1} \\
-\left( N-1\right) \int d\mathbf{%
q}_{12}\;\delta \left( q_{12}-\sigma \right) \left[ \widehat{b}_{12}-1\right]
\Theta \left( -\mathbf{q}_{12}\cdot \mathbf{p}_{12}\right) \rho
_{2}(x_{1},x_{2};t).  \notag
\end{eqnarray}%
If the two-body distribution on the right is approximated by $\rho
_{2}(x_{1},x_{2};t)\approx \rho _{1}(x_{1};t)\rho _{1}(x_{2};t)g(\mathbf{q}%
_{1},\mathbf{q}_{2};t)$, the result is the well-known Enskog equation for
the one-body distribution of a system of hard spheres (here, $g(\mathbf{q}%
_{1},\mathbf{q}_{2};t)$ is the probability to find two atoms at positions $%
\mathbf{q}_{1}$ and $\mathbf{q}_{2}$ and is normally approximated by
the equivalent local equilibrium function). In fact, examination of eq.(\ref{BBGKY1})
shows that the necessary approximation is actually
\begin{equation}
\delta \left( q_{12}-\sigma \right) \Theta \left( -\mathbf{q}_{12}\cdot 
\mathbf{p}_{12}\right) \rho _{2}(x_{1},x_{2};t)\approx \delta \left(
q_{12}-\sigma \right) \Theta \left( -\mathbf{q}_{12}\cdot \mathbf{p}%
_{12}\right) \rho _{1}(x_{1};t)\rho _{1}(x_{2};t)g_{0}(\mathbf{q}_{1},%
\mathbf{q}_{2};t)  \label{chaos}
\end{equation}%
which is a somewhat weaker approximation than the assumption that the
two-body distribution always factorizes: rather, one need only assume that
it factorizes for the case that the two atoms are in contact and approaching
one another which is to say, just prior to a collision. This is a precise
statement, for hard-spheres, of Boltzmann's ''assumption of molecular
chaos''. Immediately after a collision, the direction of the relative
momentum is reversed and the momenta of the two atoms are obviously
correlated. In fact, it has been shown\cite{LutskoHCS} that the approximation
given in Eq.(\ref{chaos}) implies the form of the entire two-body
distribution at contact is given by%
\begin{eqnarray}
\delta \left( q_{12}-\sigma \right) \rho _{2}(x_{1},x_{2};t) &\simeq &\delta
\left( q_{12}-\sigma \right) \rho _{1}(x_{1};t)\rho _{1}(x_{2};t)g_{0}(%
\mathbf{q}_{1},\mathbf{q}_{2};t)  \notag \\
&&+\delta \left( q_{12}-\sigma \right) \Theta \left( \mathbf{q}_{12}\cdot 
\mathbf{p}_{12}\right) \left[ \widehat{b}_{12}-1\right] \rho
_{1}(x_{1};t)\rho _{1}(x_{2};t)g_{0}(\mathbf{q}_{1},\mathbf{q}_{2};t).
\label{chaos1}
\end{eqnarray}%
The distribution is seen to have to parts: the first term on the right describing uncorrelated
atoms the second term describing velocity correlations
which arise because of collisions. In equilibrium, the second term vanishes
but for non-equilibrium systems, it is generally present and can give rise
to substantial structural effects as will be discussed below. This relation
is critical in that it can be used to calculate, to the same level of
approximation as is inherent in the Enskog equation, any static two-body
correlation function at contact.

Finally, the nonequilibrium pair distribution function is defined, as in
equilibrium, by%
\begin{equation}
g(\mathbf{q}_{1},\mathbf{q}_{2};t)=V\int d\mathbf{p}_{1}d\mathbf{p}_{2}\rho
_{2}(x_{1},x_{2};t).
\end{equation}%
From the definition of the local density field%
\begin{equation}
n\left( \mathbf{r}\right) =\sum_{i=1}^{N}\delta \left( \mathbf{r-q}%
_{i}\right)
\end{equation}%
it follows that the pdf is related to the density autocorrelation function
via the usual relationship%
\begin{equation}
\left\langle n\left( \mathbf{r}\right) n\left( \mathbf{r}^{\prime }\right);t
\right\rangle = n\delta\left(\mathbf{r}-\mathbf{r}^{\prime }\right)+n^{2} g\left(\mathbf{r},\mathbf{r}^{\prime };t\right)
\end{equation}
where the brackets indicate an average over the (time-dependent) two-body distribution function.

\subsection{Uniform Shear flow}

To induce shear flow, modified periodic boundary conditions are used\cite%
{LeesEdwards}. In the $x$- and $z$-directions, periodic boundaries are
applied whereas in the $y$- direction, an atom with coordinates $\mathbf{q}%
_{i}=(x_{i},y_{i},z_{i})$ and momentum $\mathbf{p}%
_{i}=(p_{xi},p_{yi},p_{zi}) $ will have images with $\mathbf{q}%
=(x_{i}+aLt,y_{i}+L,z_{i})$ and $\mathbf{p}=(p_{xi}+aL,p_{yi},p_{zi})$ where 
$t$ is the time and the parameter $a$, having the units of inverse time, is
the shear rate. These are just periodic
boundaries applied to the coordinates $\mathbf{q}_{i}^{\prime }=\mathbf{q}%
_{i}-aty_{i}\widehat{\mathbf{x}}$ and $\mathbf{p}_{i}^{\prime }=\mathbf{p}%
_{i}-ay_{i}\widehat{\mathbf{x}}$ which are the atomic coordinates in the
local rest frame of a system undergoing uniform shear flow and are the
standard means by which such a flow is simulated.

It is known\cite{LutskoEnskog} that this combination of dynamics and
boundary conditions allows for an exact solution of the macroscopic
conservation laws in which the local density is constant, the local flow
velocity is given by $\mathbf{v}(\mathbf{r})=ar_{y}\widehat{\mathbf{x}}$ and
the local temperature, defined as the excess kinetic energy relative to the
flow field, is spatially uniform in the co-moving frame and increase as%
\begin{equation}
\frac{3}{2}nk_{B}\frac{\partial }{\partial t}T=-aP_{xy}+F
\end{equation}%
where $P_{ij}$ is the macroscopic pressure tensor, which is also spatially
uniform, and the last term on the right represents the effect of the
external forces. Typically, an external force, or thermostat, is included
such that the right hand side of this equation vanishes, thus giving a
constant temperature and allowing for the possibility of a stationary state.
Here, it is assumed that for shear rates below the ordering transition, all
thermostats are equivalent\cite{DuftyThermostat,Evans}. 

The one-body distribution function of a sheared and thermostated fluid of
hard-spheres has been studied in considerable detail\cite{LutskoEnskog},\cite{Lutsko_EnskogPRL},%
and may be approximated as%
\begin{eqnarray}
f(\mathbf{q},\mathbf{p)} &\mathbf{=}&\rho \left( \frac{\beta }{2\pi }\right)
^{3/2}\left[ \det (\Delta )\right] ^{-1/2}\exp (-\frac{1}{2}\beta
p_{i}^{\prime }p_{j}^{\prime }\Delta _{ij}^{-1}) \\
\Delta _{ij} &=&\delta _{ij}+A_{ij}
\end{eqnarray}%
where the (constant) matrix of coefficients is defined implicitly as the
solution of%
\begin{equation}
a\left( \delta _{xi}\delta _{jy}+\delta _{yi}\delta _{jx}+\delta
_{xi}A_{jy}+\delta _{xj}A_{iy}\right) =C_{ij}^{(0)}+C_{ij,lm}^{(1)}A_{lm}
\label{1body}
\end{equation}%
with%
\begin{eqnarray}
C_{ij}^{(0)} &=&\rho \chi \int d\mathbf{q}\;\delta (q-1)q_{i}q_{j}\left\{ 
\frac{1}{\sqrt{\pi }}w^{2}e^{-w^{2}/4}+\frac{1}{2}(w^{2}+2)\psi (w)\right\}
\\
C_{ij,lm}^{(1)} &=&\frac{1}{2}\rho \chi \int d\mathbf{q}\;\delta (q-1)\left[
\left\{ \frac{8}{\sqrt{\pi }}e^{-w^{2}/4}+6\psi (w)\right\}
q_{i}q_{j}q_{l}q_{m}+\left\{ \frac{2}{\sqrt{\pi }}e^{-w^{2}/4}+\psi
(w)\right\} \Delta _{ijlm}\right]  \notag \\
\Delta _{ijlm} &=&\left( \delta _{il}q_{j}q_{m}+\delta
_{im}q_{j}q_{l}+\delta _{jl}q_{i}q_{m}\{\delta _{jm}q_{i}q_{l}\right)  \notag
\\
\psi (w) &=&w\left( \func{erf}\left( \frac{w}{2}\right) -1\right)  \notag \\
w &=&aq_{x}q_{y}  \notag
\end{eqnarray}%
together with the condition that $Tr(A)=0$. This approximation gives a
semi-quantitative description of effects of strong shear such as shear
thinning and normal stresses as well as being positive-definite at all shear
rates. The pdf at contact within this approximation is found to be%
\begin{equation}
\delta \left( r_{12}-\sigma \right) g(\mathbf{r}_{1},\mathbf{r}%
_{2})=\delta \left( r_{12}-\sigma \right) \chi _{0}\left( 1-\func{erf}%
\left( \frac{1}{2\sigma ^{2}}\frac{ar_{12x}r_{12y}}{\sqrt{%
1+A_{lm}r_{12l}r_{12m}}}\right) \right)  \label{contact}
\end{equation}%
which follows directly from Eq.(\ref{1body}) and (\ref{chaos1}). Here and below, $\chi_{0}$ 
is taken to be the equilibrium value of the pdf at contact as calculated in the
Carnahan-Starling approximation. From this, the
projections of the pdf at contact onto the spherical harmonics may be
calculated as%
\begin{equation}
M_{lm}\equiv \int d\mathbf{r}_{12}\;Y_{lm}^{\ast }\left( \widehat{\mathbf{r}}%
_{12}\right) \delta \left( r_{12}-\sigma \right) g(\mathbf{r}_{1},\mathbf{r}%
_{2}) . \label{contact_moments}
\end{equation}
Note that from this point, the dependence of all quantities on time is being suppressed since we work
in a steady state.

\section{The nonequilibrium pair distribution function}

The Enskog approximation gives information about quantities at
contact. In order to understand the pdf for finite separations, several
different approaches have been suggested. Two well-known proposals, the
kinetic model of Hess\cite{Hess85} and the fluctuation model of Ronis\cite{RonisShear}, involve
phenomenological parameters which can be fixed by requiring that they
reproduce one of the moments $M_{lm}$ as calculated from \ref{contact_moments}.
These theories share the property that in equilibrium, they reduce to the
equilibrium pdf so that it is reasonable to attempt to reproduce local
information such as the moments at contact. This contrasts with calculations
of the density autocorrelation function based on kinetic theory\cite%
{TK_LightScattering},\cite{Mirim} or, equivalently, long-wavelength Langevin
models\cite{Lutsko_Fluctuations} which can only give information at
asymptotically large separations for which the information at contact is not
relevant. 

\subsection{Kinetic Model}

The second equation of the BBGKY hierarchy is%
\begin{eqnarray}
&&\frac{\partial }{\partial t}\rho _{2}(x_{1},x_{2};t)+\sum_{i=1,2}\left( 
\mathbf{p}_{i}^{\prime }\cdot \frac{\partial }{\partial \mathbf{q}_{i}}+%
\mathbf{q}_{i}\cdot \overleftrightarrow{a}\cdot \frac{\partial }{\partial 
\mathbf{q}_{i}}+\frac{\partial }{\partial \mathbf{p}_{i}^{\prime }}\cdot 
\mathbf{F}(x_{i})\right) \rho _{2}(x_{1},x_{2};t)+\overline{T}_{-}(12)\rho
_{2}(x_{1},x_{2};t)  \notag \\
&=&-N\int d3\;\left( \overline{T}_{-}(13)+\overline{T}_{-}(23)\right) \rho
_{3}(x_{1},x_{2},x_{3};t)
\end{eqnarray}%
and integrating over the momenta gives an equation for the pdf%
\begin{eqnarray}
&&\frac{\partial }{\partial t}g_{2}(\mathbf{q}_{12};t)+\mathbf{q}_{12}\cdot 
\overleftrightarrow{a}\cdot \frac{\partial }{\partial \mathbf{q}_{12}}g_{2}(%
\mathbf{q}_{12};t)  \notag \\
&=&-\int d\mathbf{p}_{1}d\mathbf{p}_{2}\overline{T}_{-}(12)\rho
_{2}(x_{1},x_{2};t)-n\int d\mathbf{p}_{1}d\mathbf{p}_{2}\int d3\;\left( 
\overline{T}_{-}(13)+\overline{T}_{-}(23)\right) \rho _{3}(x_{1},x_{2};t).
\end{eqnarray}%
The kinetic models studied by Hess et al\cite{Hess85} consist in replacing
the complicated right hand side of this equation my a simpler diffusion or
relaxation model constrained only by the requirement that it force a
relaxation towards the equilibrium state. The simplest model then takes the
form%
\begin{equation}
\frac{\partial }{\partial t}g_{2}(\mathbf{q}_{12};t)+\mathbf{q}_{12}\cdot 
\overleftrightarrow{a}\cdot \frac{\partial }{\partial \mathbf{q}_{12}}g_{2}(%
\mathbf{q}_{12};t)=-\Gamma \left( g_{2}(\mathbf{q}_{12};t)-g_{2}^{eq}(%
q_{12})\right)  \label{hess1}
\end{equation}%
or, Fourier transforming,%
\begin{equation}
\frac{\partial }{\partial t}\widetilde{g}_{2}(\mathbf{k}_{12};t)-\mathbf{k}%
_{12}\cdot \overleftrightarrow{a}\cdot \frac{\partial }{\partial \mathbf{k}%
_{12}}\widetilde{g}_{2}(\mathbf{k}_{12};t)=-\Gamma \left( \widetilde{h}(\mathbf{k%
}_{12};t)-\widetilde{h}_{2}^{eq}(k_{12})\right) .  \label{hess3}
\end{equation}%
The solution to Eq.(\ref{hess1}), under the assumption of stationarity, is%
\begin{equation}
g_{2}(\mathbf{q}_{12})=\int_{0}^{\infty }d\gamma \;e^{-\gamma
}g_{2}^{eq}\left( q_{12}(a\gamma \Gamma )\right)
\end{equation}%
where%
\begin{equation}
\mathbf{q}_{12}(a\gamma )=\left( q_{12x}-a\gamma
q_{12y},q_{12y},q_{12z}\right) .
\end{equation}%
In Fourier space, this becomes the solution to Eq.(\ref{hess3})%
\begin{equation*}
\widetilde{g}_{2}(\mathbf{k}_{12})=\int_{0}^{\infty }d\gamma \;e^{-\gamma }%
\widetilde{g}_{2}^{eq}\left( k_{12}(-a\gamma \Gamma )\right)
\end{equation*}%
with%
\begin{equation}
\mathbf{k}_{12}(-a\gamma \Gamma )=\left( k_{12x},k_{12y}+a\gamma \Gamma
k_{12x},k_{12z}\right) .
\end{equation}%
As alluded to above, the relaxation time appearing in this model can be
fixed by requiring that the model reproduce one of the moments $M_{lm}.$

\subsection{Langevin Model for Density Fluctuations}

The Langevin model of Ronis\cite{RonisShear} consists of a convective-diffusion equation for
the decay of density fluctuations which, in Fourier space, appears as%
\begin{equation}
\frac{\partial }{\partial t}\delta n\left( \mathbf{k},t\right) -\mathbf{k}%
\cdot \overleftrightarrow{a}^{T}\cdot \frac{\partial }{\partial \mathbf{k}}%
\delta n\left( \mathbf{k},t\right) -D\left( k\right) k^{2}\delta n\left( 
\mathbf{k},t\right) =f\left( \mathbf{k},t\right)
\end{equation}%
where $D\left( k\right) $ is a wave-vector-dependent diffusion constant and $%
f\left( \mathbf{k},t\right) $ is a fluctuating force representing the
neglected degrees of freedom. The fluctuating force is approximated as
delta-function correlated in both wave-vector and time, with amplitude $%
D(k)S_{0}(k)k^{2}$ where $S_{0}(k)$ is the equilibrium static structure
factor. The pdf is obtained by solving for the density fluctuation as a
functional of the force and evaluation of the equal-time density-density
correlation function with the result%
\begin{equation}
\widetilde{h}(\mathbf{k}_{12})=\int_{0}^{\infty }d\gamma \;D(k(-a\gamma
))h_{0}(k(-a\gamma ))k^{2}(-a\gamma )\exp \left( -\int_{0}^{\gamma }d\gamma
^{\prime }D\left( k(-a\gamma ^{\prime })\right) k^{2}(-a\gamma ^{\prime
})\right)
\end{equation}%
so that it is seen that the particular choice for the autocorrelation of the
force leads to the correct result in equilibrium. The similarity between
this and the Hess' model is apparent and in fact the same result is obtained
if the relaxation time in the latter, Eq.(\ref{hess3}), is taken to be
wave-vector-dependent with $\Gamma (k)=D(k)k^{2}$. To close the model, Ronis
uses $D(k)=D_{0}/S_{0}(k)$ with $D_{0}$ a constant which he takes to be the equilibrium 
self-diffusion constant although in the present circumstances, it will be
fixed by the requirement that the model give the correct value of $M_{22}$.
Finally, the nonequilibrium correction can be written more explicitly by
means of an integration by parts which gives%
\begin{equation}
\widetilde{h}(\mathbf{k}_{12})-\widetilde{h}_{0}(\mathbf{k}%
_{12})=a\int_{0}^{\infty }d\gamma \;\frac{k_{x}k_{y}(-a\gamma )}{%
k(-a\gamma )}h_{0}^{\prime }(k(-a\gamma ))\exp \left( -\int_{0}^{\gamma
}d\gamma ^{\prime }D\left( k(-a\gamma ^{\prime })\right) k^{2}(-a\gamma
^{\prime })\right) \label{ronis1}
\end{equation}%
where $h_{0}^{\prime }(k)\equiv \frac{d}{dk}h_{0}(k)$. The same result has
recently been derived\cite{RPA} using a random phase approximation in the
context of a Langevin model for the atomic coordinates.

\bigskip

\subsection{Nonequilibrium GMSA}

As in equilibrium, define the direct correlation function as usual through
the Ornstein-Zernike equation%
\begin{equation}
h(\mathbf{r}_{1},\mathbf{r}_{2})=c(\mathbf{r}_{1},\mathbf{r}_{2})+\int d%
\mathbf{r}_{3}\;c(\mathbf{r}_{1},\mathbf{r}_{3})\rho (\mathbf{r}_{3})h(%
\mathbf{r}_{3},\mathbf{r}_{2})  \label{OZ}
\end{equation}%
where $h(\mathbf{r}_{1},\mathbf{r}_{2})=g(\mathbf{r}_{1},\mathbf{r}_{2})-1$
is the structure function. The pdf must satisfy the boundary condition
that the probability for two atoms to interpenetrate is zero so that%
\begin{equation}
h(\mathbf{r}_{1},\mathbf{r}_{2})=-1\;\;\;\left| \mathbf{r}_{1}-\mathbf{r}%
_{2}\right| <\sigma  \label{bc1}
\end{equation}%
The OZ equation can be solved for both the structure function and the direct
correlation function provided that this is supplemented by a closure
relation between the two.

In equilibrium, the relation between the pdf and the direct correlation
function may be written as%
\begin{equation}
c(\mathbf{r}_{1},\mathbf{r}_{2})=\ln g(\mathbf{r}_{1},\mathbf{r}_{2})-h(%
\mathbf{r}_{1},\mathbf{r}_{2})+v(\mathbf{r}_{1},\mathbf{r}_{2})+B(\mathbf{r}%
_{1},\mathbf{r}_{2})  \label{dcf}
\end{equation}%
where $v(\mathbf{r}_{1},\mathbf{r}_{2})$ is the pair potential and $B(%
\mathbf{r}_{1},\mathbf{r}_{2})$ is the bridge function which is
not generally known in closed form. The integral equation approach to liquid state structure
can be written in terms of various approximations to the bridge function.
Setting $B=0$  yields the hyper-netted chain equation and
further approximating $\ln g=\ln \left[ 1+h\right] \approx h$,
 or $B=h-ln\left(g\right)$, yields the Percus-Yevik approximation. A
 number of other approximations exist, 
including  schemes such as that of Rogers and Young\cite{RY} and
of Zerah and Hansen\cite{Zarah}, which involve a parameterization of the right
hand side of Eq.(\ref{dcf}). 
For hard-core potentials, the Percus-Yevik approximation reduces to the statement 
$c(\mathbf{r}_{1},\mathbf{r}_{2})=0$ for $\left| \mathbf{r}_{1}-\mathbf{r}_{2}\right| >\sigma $
and the GMSA replaces the right hand side by a Yukawa function with parameters adjusted to 
give a known equation of state. In this case, these have been shown to be the first two steps
 in a systematic expansion of the tail of 
the dcf with little underlying physical approximation\cite{Yuste91,Yuste93,Yuste94,CC}. It thus becomes 
natural to carry over this model
to the nonequilibrium state so that the closure condition is expressed in
terms of a similar parameterization of the tail of the dcf giving the form%
\begin{equation}
c(\mathbf{r}_{1},\mathbf{r}_{2})=\sum_{i}A_{i}K_{i}(\mathbf{r}_{1},\mathbf{r}%
_{2})\;\;\;\left| \mathbf{r}_{1}-\mathbf{r}_{2}\right| >\sigma  \label{bc2}
\end{equation}%
for some set of basis functions $\left\{ K_{i}(\mathbf{r}_{1},\mathbf{r}%
_{2})\right\} $. As it stands, Eq.(\ref{bc2}) is quite general and the
physical approximation will be to truncate and parameterize this expansion
as discussed below.

The problem of solving the OZ equation is now formally equivalent to that of
the case of molecular fluids and similar techniques can be used\cite{Gray}.
To begin, one expands the angular dependence of the dcf, the pdf and the
boundary conditions in terms of spherical harmonics so that%
\begin{equation}
h({}\mathbf{r}_{1},\mathbf{r}_{2};t)=\sum_{lm}h_{lm}(r_{12};t)Y_{lm}(%
\widehat{r}_{12})
\end{equation}%
with similar expansions for the other quantities. Using Rayleigh's expansion
of a plane wave in terms of radial and angular functions and the addition
theorem for spherical harmonics, it is easy to show\cite{Gray} that the Fourier
transform of such an expansion has the form%
\begin{equation}
\widetilde{h}(\mathbf{k}_{1},\mathbf{k}_{2};t)=\left( 2\pi \right)
^{3}\delta \left( \mathbf{k}_{1}+\mathbf{k}_{2}\right) \sum_{lm}\overline{h}%
_{lm}(k_{1};t)Y_{lm}(\widehat{\mathbf{k}}_{1})
\end{equation}%
with the coefficients defined in terms of Hankel transforms 
\begin{equation}
\overline{h}_{lm}(k;t)=4\pi i^{l}\int_{0}^{\infty }r^{2}dr\mathbf{\;}%
j_{l}(kr)h_{lm}\left( r\right).
\end{equation}%
The Fourier transform of the OZ equation then becomes
\begin{eqnarray}
\overline{h}_{lm}\left( k\right) &=&\overline{c}_{lm}\left( k\right) +n%
\overline{c}_{l^{\prime }m^{\prime }}\left( k\right) \overline{h}_{l^{\prime
\prime }m^{\prime \prime }}\left( k\right) \int d\widehat{\mathbf{k}}%
\;Y_{l^{\prime }m^{\prime }}(\mathbf{k})Y_{l^{\prime \prime }m^{\prime
\prime }}(-\mathbf{k})Y_{lm}^{\ast }(\mathbf{k})  \label{final} \\
&=&\overline{c}_{lm}\left( k\right) +n\frac{1}{\sqrt{4\pi }}\sum_{\left|
l^{\prime }-l^{\prime \prime }\right| \leq l\leq l^{\prime }+l^{\prime
\prime }}\sum_{m^{\prime }=-l^{\prime }}^{l^{\prime }}A(l,l^{\prime
},l^{\prime \prime },m,m^{\prime })\overline{c}_{l^{\prime }m^{\prime
}}\left( k\right) \overline{h}_{l^{\prime \prime }m-m^{\prime }}\left(
k\right)  \notag
\end{eqnarray}%
where%
\begin{equation}
A(l,l^{\prime },l^{\prime \prime },m,m^{\prime })=(-1)^{l^{\prime }+2m}\sqrt{%
\frac{\left( 2l^{\prime }+1\right) \left( 2l^{\prime \prime }+1\right) }{2l+1%
}}C(l^{\prime },l^{\prime \prime },l|000)C(l^{\prime },l^{\prime \prime
},l|m^{\prime },m-m^{\prime },m)  \label{coeff}
\end{equation}%
and the last line is a well-known result which follows from the
Wigner-Eckart theorem\cite{ArfkenWeber}. Equation (\ref{final}) together with Eq.(\ref{bc2})
and the exact condition, Eq.(\ref{bc1}) serve to define the integral
equation.

In the theory of molecular liquids\cite{Gray}, auxiliary functions are
usually defined as%
\begin{equation}
f_{lm}^{\prime }(r)=\left\{ 
\begin{array}{c}
\frac{4\pi i}{\left( 2\pi \right) ^{3}}\int_{0}^{\infty }k^{2}dk\;\frac{\sin
\left( kr\right) }{kr}\overline{f}_{lm}\left( k\right) \;\qquad l\;\text{even%
} \\ 
\frac{4\pi i}{\left( 2\pi \right) ^{3}}\int_{0}^{\infty }k^{2}dk\;\frac{\cos
\left( kr\right) }{kr}\overline{f}_{lm}\left( k\right) \;\qquad l\;\text{odd}%
\end{array}%
\right.
\end{equation}%
where $\overline{f}_{lm}(k)$ could be either $\overline{h}_{lm}(k)$ or $%
\overline{c}_{lm}(k)$. For even values of $l$, the only ones of concern
below, this means that the Hankel-transform of the original functions is the
Fourier-transform of the auxiliary function. This is useful for numerical
work but more important is that the auxiliary functions tend to be of
shorter range than the original functions. To see this, we need the relation
between the auxiliary functions and original functions in real space%
\begin{equation}
f_{lm}^{\prime }(r)=f_{lm}(r)-\int_{r}^{\infty }r^{\prime 2}dr^{\prime }\;%
\frac{1}{rr^{\prime 2}}P_{l}^{\prime }\left( r/r^{\prime }\right)
f_{lm}(r^{\prime })  \label{aux}
\end{equation}%
where $P_{l}^{\prime }(u)$ is the derivative of the $l-$th Legendre
polynomial with respect to its argument. Direct calculation shows that if $%
f_{lm}(r)=r^{-n}$ then $f_{lm}^{\prime }(r)=0$ provided that $2<n<l+3$ .
This transformation therefore removes a subset of long-ranged decays and is consequently degenerate. The
inverse transformation is%
\begin{equation}
f_{lm}(r)=\left( -1\right) ^{l}f_{lm}^{\prime }(r)-\left( -1\right)
^{l}\int_{0}^{r}r^{\prime 2}dr^{\prime }\;\frac{1}{r^{2}r^{\prime }}%
P_{l}^{\prime }\left( r^{\prime }/r\right) f_{lm}^{\prime }(r^{\prime
})+\sum_{n=1}^{\left[ \frac{l-1}{2}\right] }A_{lm,n}r^{-(2n+1)}.
\label{uniqueness}
\end{equation}%
where the relationship holds for any choice of the coefficients $A_{lm,n}$ - an expression
of the degeneracy of the original transformation. In the present application,  since $h_{lm}\left( 0\right) $ 
is, from eq.(\ref{bc1}), we will always have $A_{lm,n}=0$ whereas for the dcf, no such statement can be made.

Further discussion of the details of the solution of these equations is
given in appendix \ref{App:Exact} and only some of the
conclusions of that analysis are stated here. First, because of the symmetry of the boundary
conditions, only coefficients corresponding to even values of $l$ and $m$
will be nonzero. Second, one expects that, because of the symmetry of the
flow, the dominant nonequilibrium contributions come from $l=2$ and $m=\pm 2$
(since $Y_{22}+Y_{2-2}\propto \widehat{x}\widehat{y}$) and indeed this can
be verified for the pdf at contact by calculations using Eq.(\ref{contact}).
In the case that we keep only these contributions to the OZ equation, as
well as the $l=m=0$ component necessary to describe the equilibrium
contribution, the problem can be transformed into the solution of two one-dimensional OZ
equations with the Yukawa closure and an analytic solution is possible as described in appendix \ref{App:Exact}%
. This approximation should be understood in the spirit of a truncation of a
moment solution rather than an expansion in the shear rate and in fact, all
of the expressions presented below depend \emph{nonlinearly} on the shear
rate. The result is that 
\begin{eqnarray}
h(\mathbf{r}) &=&h_{00}(r)Y_{00}(r)+2h_{22,r}(r)\func{Re}Y_{22}(\widehat{\mathbf{r}}%
)-2h_{22,i}(r)\func{Im}Y_{22}(\widehat{\mathbf{r}})  \label{model1} \\
&=&\frac{1}{\sqrt{4\pi }}h_{00}(r)+\sqrt{\frac{5}{8\pi }}\func{Re}\left(
h_{22}(r)\right) \left( \widehat{r}_{x}^{2}-\widehat{r}_{y}^{2}\right) -%
\sqrt{\frac{5}{2\pi }}\func{Im}\left( h_{22}(r)\right) \widehat{r}_{x}%
\widehat{r}_{y}  \notag
\end{eqnarray}%
where the angular average of the pdf is given by

\begin{equation}
h_{00}(r)=\frac{\sqrt{4\pi }}{2n}\left(
n_{+}h_{+}(r;n_{+})+n_{-}h_{-}(r;n_{-})\right)  \label{model2}
\end{equation}%
with $n_{\pm }=n\left[ 1\pm \left| c\right| \sqrt{\frac{1}{4\pi }}\func{Im}%
\left( B_{22,1}\right) \right] $ and where the constant, $B_{22,1}$, is
related to $A_{22,1}$ occuring in eq(\ref{uniqueness}). As discussed in detail in the
appendix, the functions $h_{\pm }(r;n)$ may be expressed in terms of the
solution of the OZ equation for \emph{homogeneous} system (e.g., the
Percus-Yevik solution if the right hand side of Eq.(\ref{bc2}) is set to
zero or the known analytic solution of the N-Yukawa closure\cite{Hoye77} if
the basis functions are Yukawas) for density $n_{\pm }$. The anisotropic
component is given by%
\begin{eqnarray}
h_{22}(r) &=&\Theta \left( r-1\right) \left[ h_{22}^{\prime }\left(
r^{\prime };t\right) +\frac{B_{22,1}}{r^{3}}-\frac{3}{r^{3}}%
\int_{1}^{r}r^{\prime 2}dr^{\prime }\;h_{22}^{\prime }\left( r^{\prime
};B_{22,1}\right) \right]  \label{model3} \\
\func{Im}\left( h_{22}^{\prime }(r;B_{22,1})\right) &=&\frac{\sqrt{4\pi }}{%
2\left| a\right| n}\left( n_{+}h_{+}(r;n_{+},)-n_{-}h_{-}(r;n_{-})\right) 
\notag \\
\func{Re}\left( h_{22}^{\prime }(r)\right) &=&\func{Im}\left( h_{22}^{\prime
}(r)\right) \func{Re}\left( M_{22}\right) /\func{Im}\left( M_{22}\right) 
\notag
\end{eqnarray}%
Aside from the truncation of the OZ
hierarchy, these results are independent of any assumption about the closure
condition given in Eq.(\ref{bc2}).

In equilibrium, the GMSA is based on the choice of a Yukawa function for the
tail of the dcf. This is motivated by the expectation that the tail is
short-ranged and, then, because a Yukawa closure is analytically tractable.
As discussed above, recent work has shown that the Yukawa may be thought of
as the first term in a systematic expansion thus removing some of the
arbitrarity of this choice. In the same spirit, we therefore take as the principle hypothesis
of the extension of the GMSA to nonequilibrium systems that the tail of
the auxiliary dcf function can be expanded as%
\begin{equation}
\Theta \left( r_{12}-\sigma \right) c^{\prime }(\mathbf{r}_{1},\mathbf{r}%
_{2})=\sum_{l,m}v_{lm}^{\prime }(r_{12})Y_{lm}\left( \widehat{\mathbf{r}}%
_{12}\right)  \label{closure1}
\end{equation}%
with
\begin{equation}
v_{lm}^{\prime }(r)\simeq \frac{1}{r}K_{lm}\exp (-z_{lm}r)\;\;\text{.}
\label{closure}
\end{equation}%
For the spherically symmetric component $l=m=0$ this is just the Yukawa
closure as in equilibrium. Because of the non-uniqueness of the relation
between the dcf and the auxiliary dcf, see the discussion after Eq.(\ref%
{uniqueness}), this corresponds to a closure of the actual dcf of the form%
\begin{equation}
\Theta \left( r_{12}-\sigma \right) c_{lm}(\mathbf{r}_{1},\mathbf{r}%
_{2})=v_{lm}(r_{12})+\sum_{n=1}^{\left[ \frac{l-1}{2}\right]
}A_{lm,n}r^{-(2n+1)}.  \label{closure3}
\end{equation}%
Since the dcf is of no significance in the present context, the values of $A_{lm,n}$ are 
left indeterminate at this stage.

As formulated, the truncated analytic model has 5 parameters corresponding
to the amplitudes and length scales of the two Yukawa terms and the constant $B_{22,1}$
appearing in Eqs.(\ref{model1})-(\ref{model3}). These parameters are constrained by two
boundary conditions consisting of the values of $M_{00}$ and $M_{22}$. In
the first application to shear flow\cite{Lutsko2001}, the model was
simplified by setting $K_{00}=K_{22}=0$. The justification for this was
simplicity, since there is then only the non-uniqueness parameter, $B_{22,1}$%
, and it was shown that a value could be found which simultaneously
satisfied both boundary conditions reasonably well. However, recent estimates%
\cite{LutskoDufty-LongRangeShear} indicate that $h_{22}\left( r\right) $
decays faster than $1/r^{3}$ for large $r$ leading to the condition that the
inverse-cube terms in Eq.(\ref{model3}) vanish in the limit $r \rightarrow \infty$ giving the constraint 
\begin{equation}
B_{22,1}=3\int_{1}^{\infty }r^{\prime 2}dr^{\prime }\;h_{22}^{\prime }\left(
r^{\prime };B_{22,1}\right) .  \label{Beq}
\end{equation}%
This still leaves two parameters undetermined. One of these will be
eliminated by taking the length scale of the length scale of the $%
v_{00}^{\prime }(r)$ function to be fixed at its equilibrium value and to
only allow the amplitude to be adjusted so as to reproduce $M_{00}$. This
still leaves one undetermined parameter which can be taken to be $z_{22}$.
In an application of this approach to granular fluids, a similar
indeterminacy was solved by insisting that the compressibility equation
continue to hold in the nonequilibrium state. Here, this is not useful
because $v_{22}(r)$ has little influence on $h_{00}$ which would be the
object fixed by such a relation. (In fact, this could be used as an
alternate means of fixing $z_{00}$.) With no other exact or
well-approximated property to fit, it seems appropriate to try to minimize the 
perturbation of the tail of the dcf. The tail of the full dcf is found to be%
\begin{equation}
c_{22}(r)=K_{22}\frac{\exp \left( -z_{22}\left( r-1\right) \right) }{r}\frac{%
r^{2}z_{22}^{2}-3z_{22}r-3}{r^{2}z_{22}^{2}}-\frac{3}{r^{3}}\int_{0}^{\infty
}c_{22}(x)x^{2}dx\;\;\;\;\text{for }r>\sigma 
\end{equation}%
which is clearly short ranged if the last term on the right vanishes
as it in fact does in the present approximation as a result of the condition given in
Eq.(\ref{Beq}). For large separations, the tail is therefore the
same as that of the auxiliary function, a Yukawa and this cannot be changed
by any condition on $z_{22}$. At short range, the Yukawa is modified and one
possibility that suggests itself is to demand that at contact, the tail
assume its equilibrium value - namely, zero. This implies that $z_{22}=\frac{%
3}{2}+\frac{1}{2}\sqrt{21}\simeq \allowbreak 3.\,\allowbreak 8$ which is the
value used below. 

\section{Molecular Dynamics Simulations}

In order to evaluate the model for the structure proposed above, I have
performed molecular dynamics simulations of sheared hard spheres in three
dimensions. The simulation makes use of Lees-Edwards boundary conditions to
impose the shear and the heating is controlled by periodically rescaling the
velocities. Specifically, to maintain an average temperature $T_{0}$, the
velocities are rescaled to give an instantaneous temperature of $0.95T_{0}$
whenever the instantaneous temperature exceeds $1.05T_{0}$. The simulations
reported here were performed using 500 atoms except where noted
below. In all cases, a cubic simulation cell was used. The equilibration
procedure consisted of first creating an equilibrium liquid at the
desired density. After $10^{7}$ collisions, the shear rate was then
instantaneously set to the desired value and the system allowed to relax
under the Lees-Edwards boundary conditions for another $10^{7}$ collisions.
Finally, the simulation was extended for another $10^{7}$ collisions during
which statistical averages were accumulated under the ergodic hypothesis. In
order to estimate the accuracy of the quantities obtained, Erpenbeck's
pooling method\cite{Erpenbeck83} was used whereby averages were accumulated
over periods of $10^{5}$ collisions and stored. The reported values were
subsequently computed by averaging these partial-averages and the standard
error of these partial averages, i.e. the standard deviation divided by the
square root of the number of observations, used as an estimate of their
accuracy. Except as noted below, the error bars in all figures are smaller than
the symbols used to display the data. Finally, all simulations were performed for reduced densities of $%
n^{\ast }\equiv n\sigma ^{3}=0.1,0.25,0.5$ and $0.75$. Based on the difference
of the equilibrium pdf at contact from its low density value, namely $\chi
_{0}=1$, these densities correspond to low density ($\chi _{0}=1.14$), moderately dense ($\chi _{0}=1.4$),
dense ($\chi _{0}=2.15$) and very dense ($\chi _{0}=3$) fluids respectively. Shear rates are reported
in units of the Boltzmann collision time $a^{\ast} = a/\left(4n^{\ast}\sqrt{\pi k_{B} T}\right)$.

\subsection{The pair distribution at contact}

\begin{figure*}
\includegraphics[angle=0,scale=0.4]{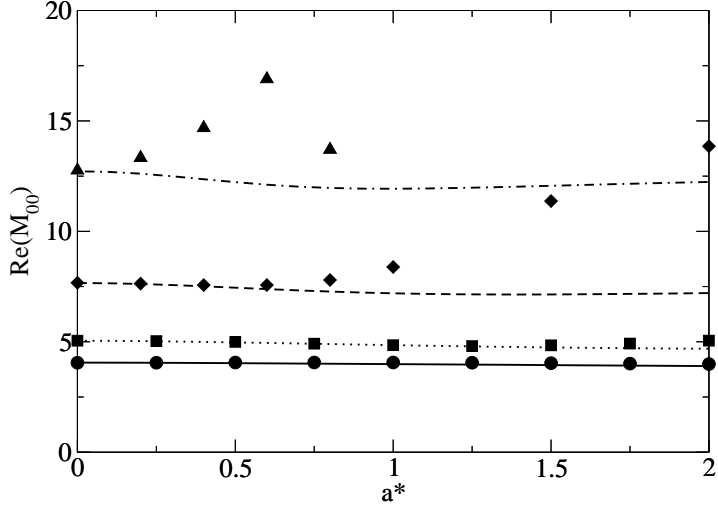}
\caption{\label{fig1} $Re\left(M_{00}\right)$ as a function of the reduced shear rate, $a^{*}$ for densities
of $0.1$ (circles), $0.25$ (squares), $0.5$ (diamonds) and $0.75$ (triangles). The
lines are the predicted values calculated from eq.(\ref{contact}).Note the non-monotonic behavior at the highest density.}
\end{figure*}

\begin{figure*}
\includegraphics[angle=0,scale=0.4]{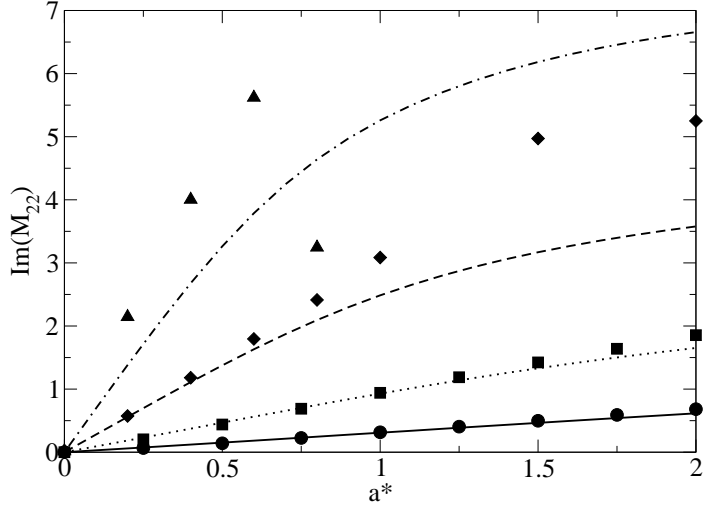}
\caption{\label{fig2} Same as Fig. \ref{fig1} but showing $M_{22}$. }
\end{figure*}

\begin{figure*}
\includegraphics[angle=0,scale=0.4]{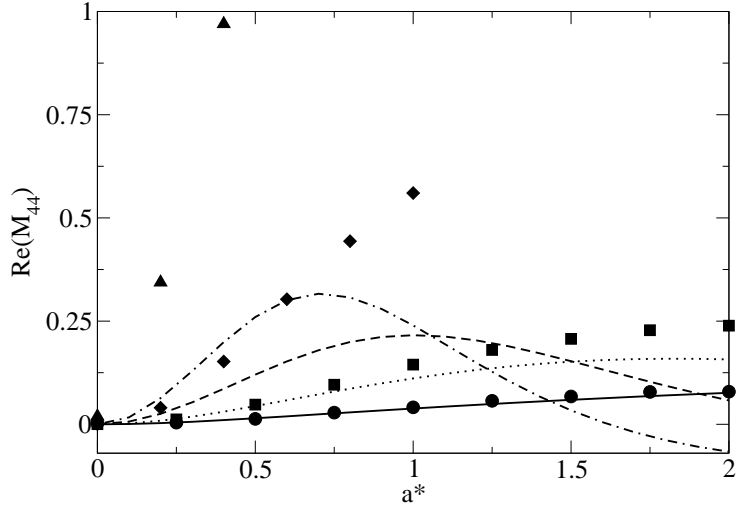}
\caption{\label{fig3} Same as Fig. \ref{fig1} but showing $M_{44}$.}
\end{figure*}

Figures \ref{fig1},\ref{fig2} and \ref{fig3} show the projections $M_{lm}$ of the pdf at contact onto
the spherical harmonics for $lm=00,22$ and $44$ as a function of shear rate
which accounts for angular dependencies of the form $1$, $\widehat{q}_{x}%
\widehat{q}_{y}$ and $\widehat{q}_{x}^{2}-\widehat{q}_{y}^{2}$ respectively.
The spherical average of the pdf, $g_{00}$, shows little variation with
shear rate for $n^{\ast }=0.1 $ and $0.25$ and only begins to show significant
variation above $a^{\ast }=0.5$ for $n^{\ast }=0.5$ while at the highest
density, significant variation is observed for all shear rates and, unlike at lower density, the curve is not monotonic. In all 
cases, the Enskog prediction is a slight decrease with increasing shear rate
which is confirmed in the low density data. The system at intermediate
density is consistent with the model for small shear rates but shows an 
\emph{increase} at higher shear rates as does the high density system at all
shear rates, in qualitative disagreement with the model. The major
nonequilibrium contribution to the structure resides in the $Im\left(g_{2 2}\right)$
components which show qualitatively similar behavior: agreement with the
model at low density and all shear rates and for low shear rates at moderate
density with significant disagreement at moderate density and high shear
rates and at all shear rates at high density. The next largest
nonequilibrium contribution, $Re\left(g_{44}\right)$, is seen to be nearly an
order of magnitude smaller than $Im\left(g_{2 2}\right)$ and it is also poorly
described by the model.

\begin{figure*}
\includegraphics[angle=0,scale=0.4]{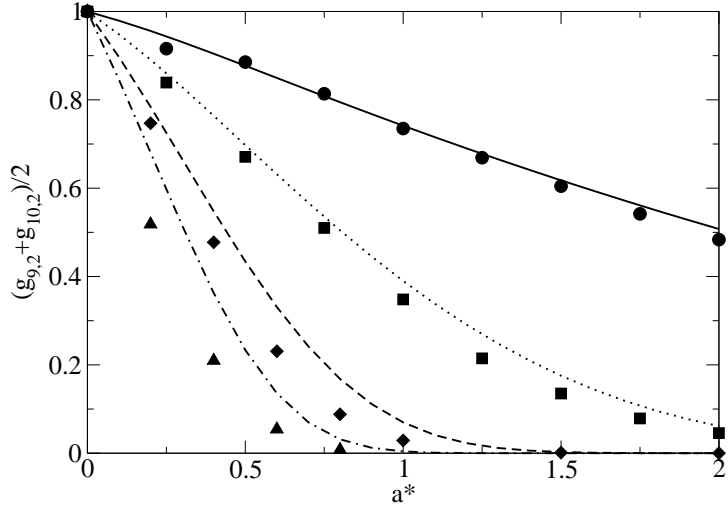}
\caption{\label{fig4} The value of $\frac{1}{2}\left(g_{9,2} + g_{10,2}\right)$ as defined in eq.\ref{glm}. The labeling is the same as in Fig. \ref{fig1}.}
\end{figure*}

\begin{figure*}
\includegraphics[angle=0,scale=0.4]{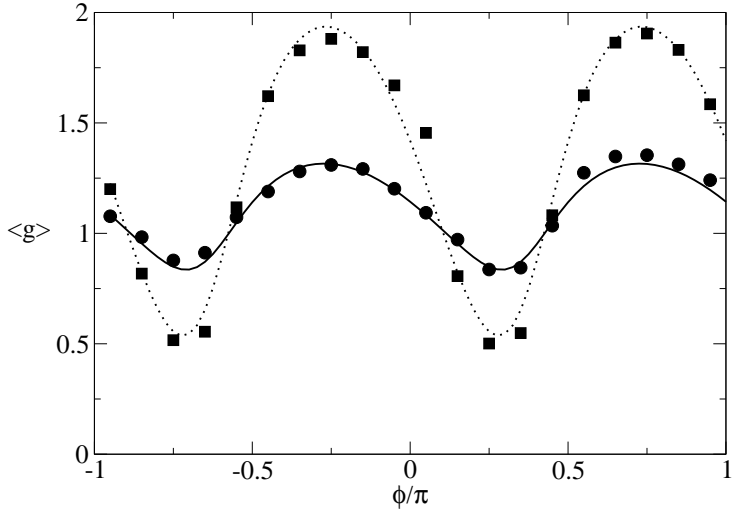}
\caption{\label{fig5} The pdf at contact averaged over a solid angle defined by
$-0.1<x<0.1$ and $\frac{m}{10} < \frac{\phi}{\pi} < \frac{m+1}{10}$ for $-10 \le m \le 10$. 
The shear rate is fixed at $a^{*}=1.0$ and the information for $n^{*}=0.1,0.25$ is labeled as in Fig.\ref{fig1}.}
\end{figure*}

\begin{figure*}
\includegraphics[angle=0,scale=0.4]{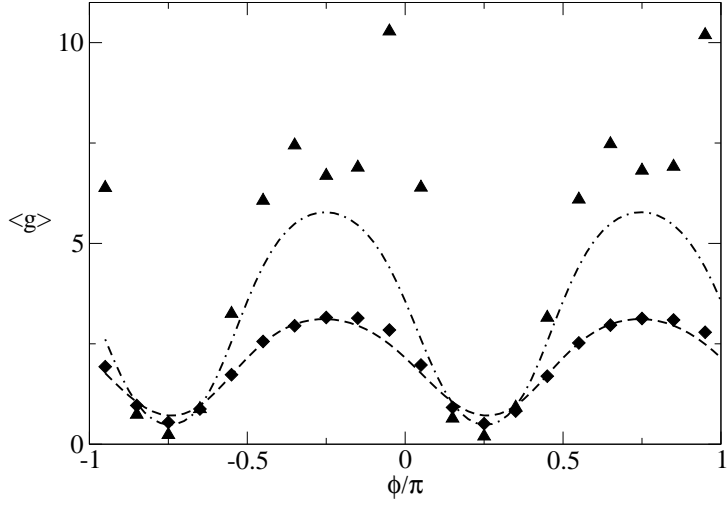}
\caption{\label{fig6} The same as Fig. \ref{fig5} for $a^{*}=0.6$ and $n^{*}=0.5,0.75$.}
\end{figure*}

\begin{figure*}
\includegraphics[angle=0,scale=0.4]{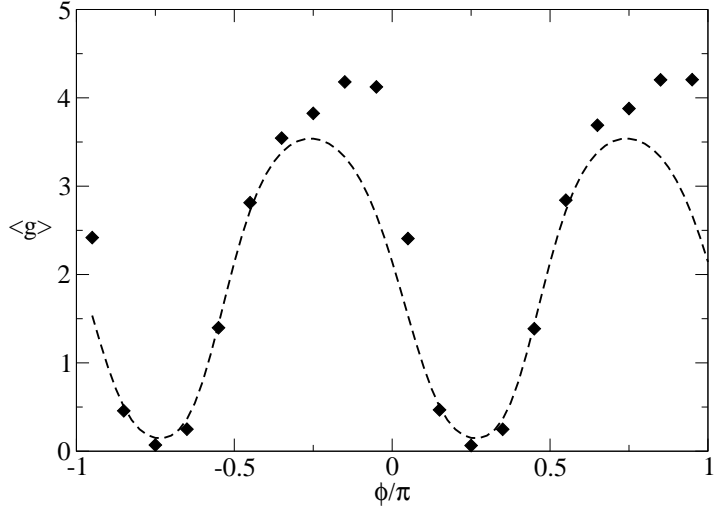}
\caption{\label{fig7} The same as Fig. \ref{fig5} for $a^{*}=1.0$ and $n^{*}=0.5$.}
\end{figure*}

These results show that the largest nonequilibrium contributions to the pdf
occur in the four directions $(\pm \frac{1}{\sqrt{2}},\pm \frac{1}{\sqrt{2}}%
,0)$. To give a direct overview of the accuracy of the models, the average
of the pdf at contact over a number of angular bins, defined as
\begin{equation}
g_{mn}=\int_{-1+m\delta _{x}}^{-1+(m+1)\delta _{x}}d\cos \theta \int_{n\delta
_{\phi }}^{\left( n+1\right) \delta _{\phi }}d\phi \;g\left( \widehat{%
\mathbf{q}}\right), \label{glm}
\end{equation}%
was monitored during the simulations for $\delta _{x}=\frac{1}{10}$ and $\delta _{\phi }=\frac{\pi }{10}$.
Figure \ref{fig4}
shows a comparison between theory and simulation of $\frac{1}{2}\left( g_{9,2}+g_{10,2}\right)$, e.g.
of the pair distribution at contact averaged over the area $-0.1 < x < 0.1$ and 
$0.2\pi < \phi < 0.3\pi$, for the various densities. This patch is centered on the direction
 $(\frac{1}{\sqrt{2}},\frac{1}{\sqrt{2}},0)$ for which the deviations from 
equilibrium are largest. The most striking feature of these
results is that the pdf drops with increasing shear until it becomes
vanishingly small indicating that no collisions take place in that
direction. The figure also shows the predicted values based on the
generalized assumption of molecular chaos, Eq.(\ref{contact_moments}),
averaged (numerically) over the same solid angle. It is evident that the
model works quite well at the lowest densities, is reasonable at the
intermediate density and is only qualitatively correct at the highest
density. In order to visualize the full directional variation of the pdf at
contact, Figure \ref{fig5} shows the spatial variation of the pdf averaged over the same
sized solid angle for the whole range of values of $\phi$ from $-\pi$ to $\pi$ for fixed 
shear rates of $1.0$ for $n^{*}=0.1,0.25$ and Fig. \ref{fig6} shows the same for $a^{*} = 0.6$ and
 $n^{*}= 0.5,0.75$. (The reason for choosing a lower shear for the higher densities will become
apparent below.) For all but the highest densities, the spatial variation is
consistent with the model, Eq.(\ref{contact}), whereas for $n^{*}=0.75$
the agreement is poor. Indeed, the simulation data in the latter case is erratic and
appears to be a superposition of a periodic function with spikes near $\phi =0$
and $\phi =\pi $ which corresponds to the directions parallel and
antiparallel to the flow (i.e., $\pm \hat{x}$). Figure \ref{fig7} shows the same quantity for $n^{\ast
}=0.5 $ and $a^{\ast }=1.0$ and the same superposition of features is
apparent. There are three possible causes for deviations from the model :
shear-induced ordering, inaccuracy of the one-body distribution and
breakdown of the assumption of molecular chaos. The structural anomalies at
high shear rate and high density suggest the former.

\bigskip

\begin{figure*}
\includegraphics[angle=0,scale=0.4]{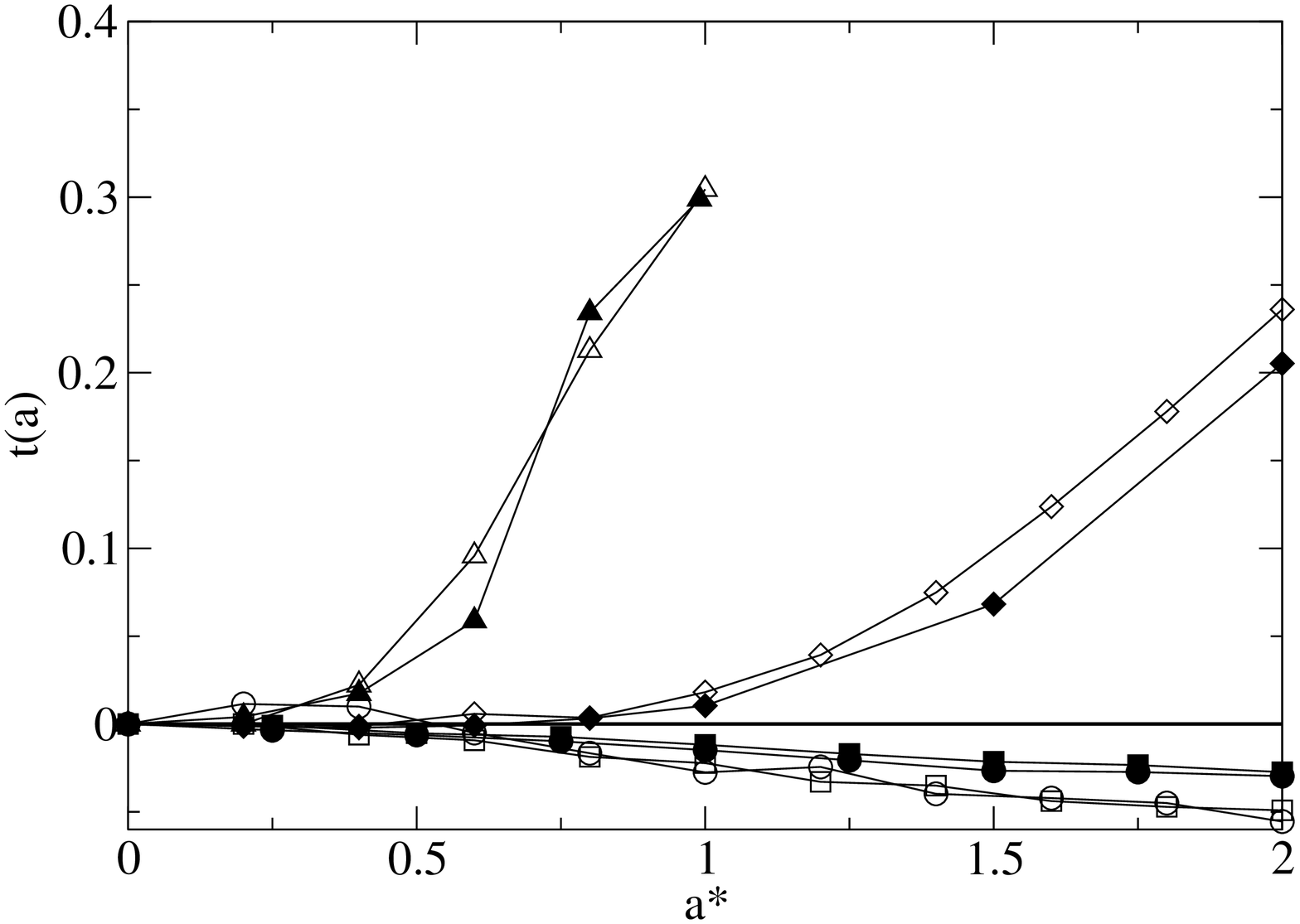}
\caption{\label{fig8} The tube density $t(a)$ as a function of the shear
rate. The labeling is the same as in Fig. \ref{fig1} except that here the lines are 
only a guide to the eye with the baseline, $t(a)=0$ indicated by the thick line. Open symbols are from simulations of 108 atoms.}
\end{figure*}

It has been known for some time that hard spheres undergo an ordering
transition at high shear rates\cite{ErpShear}. The nature of the ordered
phase remains uncertain and appears to depend on the type of thermostat used%
\cite{Evans}. For the simple rescaling thermostat used here, an ordering first
into plane perpendicular to the direction of the gradient  (here, the
y-direction) and then into strings oriented along the direction of flow, and
in a hexagonal pattern in the plane perpendicular to the flow, has been
reported\cite{ErpShear}. As a quantitative measure of such an ordering, the
average density in a tube, oriented along the direction of flow has been
monitored\cite{Lutsko96}. This is defined as%
\begin{equation}
t(u)=\frac{1}{n\pi u^{2}L}\left\langle \frac{1}{N}\sum_{i\neq j}\Theta
\left( u^{2}-q_{ij,y}^{2}-q_{ij,z}^{2}\right) \right\rangle
\end{equation}%
which can be written in terms of the pdf as 
\begin{eqnarray}
t(u) &=&\frac{1}{\pi u^{2}L}\int d\mathbf{r}\;g(\mathbf{r})\Theta \left(
u^{2}-y^{2}-z^{2}\right) \\
&=&\frac{L-2}{L}+\frac{2}{3L}u^{4}+\frac{1}{\pi u^{2}L}\int d\mathbf{r}\;h(%
\mathbf{r})\Theta \left( r^{2}-1\right) \Theta \left(
u^{2}-y^{2}-z^{2}\right) \notag
\end{eqnarray}%
and gives the average density, relative to the bulk, observed along a tube
of length $L$ and radius $u$ centered on an atom. In the limit of large $L$,
the last term on the right will only give a non-zero contribution if
long-range correlations in the direction of the flow are present (as they
would be for a ''string phase'') so that any deviation from the equilibrium
value could be attributed to the formation of such correlations. However,
for the small systems considered here, the last term will give a nonzero
contribution in all circumstances and variations of the tube density with
the shear rate could be due to variations in the pdf which nevertheless do
not involve long-ranged correlations. Figure \ref{fig8} therefore shows the tube
density as a function of shear rate for systems of both 108 atoms and 500
atoms (giving $L=6$ and $10$ respectively). For the lowest densities, the tube density actually decreases with
increasing shear rate with the decrease being larger for the smaller system. Noting that 
the size of the effect is roughly in inverse proportion to the length of the systems and independent of the density, this 
would appear to be primarily a finite size effect leading to the conclusion that in neither case is there evidence of
shear-induced changes in the density in the infinite system limit. For  $n^{\ast }=0.5$, the tube density is roughly
constant until above $a^{\ast }=0.6$ at which point it begins to rise
steadily for both system sizes. Although the magnitude of the increase is also a function of system size, the relative variation
between the two systems is much less than the over-all increase leading to the conclusion that
the increase is due to the development of long-range order.  This is consistent with our previously reported results indicating
that shear-induced ordering takes place at this point\cite{Lutsko96}. Finally, at high
density, the tube density increases dramatically above $a^{\ast
}=0.2$ indicating an ordering transition at that point. This behavior, in this case nearly independent of system size, supports the
conclusion that the deviation of the data from the molecular chaos
hypothesis is due to shear-induced ordering for $n^{\ast }=0.5$ and $a^{\ast
}>0.6$ and for virtually all of the data for the high density system. It is
also consistent with the structural data which show spikes in the pdf
corresponding to an increase in collisions in the direction of the flow.

We are then only left with the poor agreement of the model for the $g_{44}$
to explain. It seems likely that this is simply due to the inadequacy of the
information supplied for the one-body distribution. Since the distribution
is accurate only up to second moments of the velocity, it is reasonable that
the calculation is only accurate up to second order in the unit vectors.
Since these contributions are in any case small compared to the dominant $%
g_{22}$ terms in the region of validity of the model, this aspect of the
problem has not been pursued further.

\subsection{The pair distribution function at finite separations}

Here, attention is restricted to the domain of densities and shear rates
below the ordering transition. In general, the components $g_{lm}(r)$ were
estimated during the simulations by evaluating%
\begin{equation}
g_{lm}(r)=\frac{1}{N_{samples}}\sum_{samples}\frac{1}{4\pi r^{2}dr}%
\sum_{i<j}\Theta \left( r_{ij}-r\right) \Theta \left( r+dr-r_{ij}\right)
Y_{lm}^{\ast }\left( \widehat{r}_{ij}\right)
\end{equation}%
where the inner sum is an estimate of $g_{lm}(r)$ based on a snapshot of the
system and the outer sum indicates an average over many different snapshots.
The results presented here are based on snapshots taken every 10,000
collisions. Ideally, one would like to replace the outer sum by a continuous
time-average in the spirit of the ergodic theorem but the computational
expense would be prohibitive.

\begin{figure*}
\includegraphics[angle=0,scale=0.4]{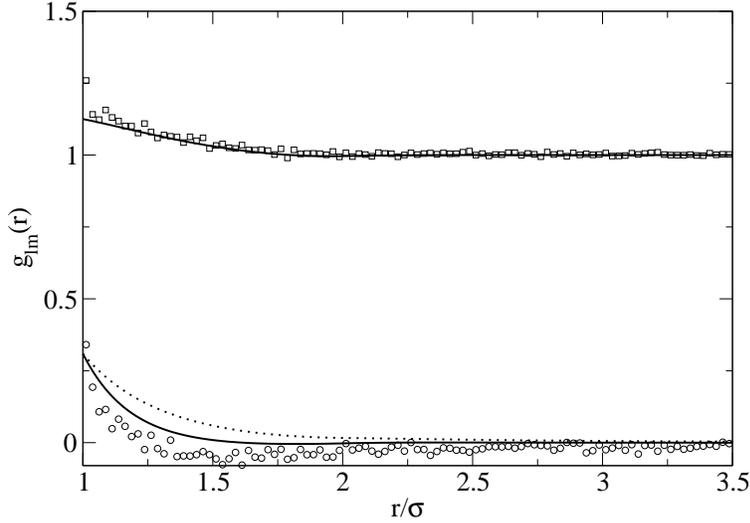}
\caption{\label{fig9} The functions $Re\left(g_{00}(r)\right)/\sqrt{4\pi}$, upper curves, and 
$Im\left(g_{22}(r)\right)$, lower curves, for $n^{*}=0.1$ and $a^{*}=1.0$ from simulation(circles), the GMSA with
the Yukawa closure (full lines) and the GMSA with the tail of the dcf set to zero 
(dotted lines).}
\end{figure*}

\begin{figure*}
\includegraphics[angle=0,scale=0.4]{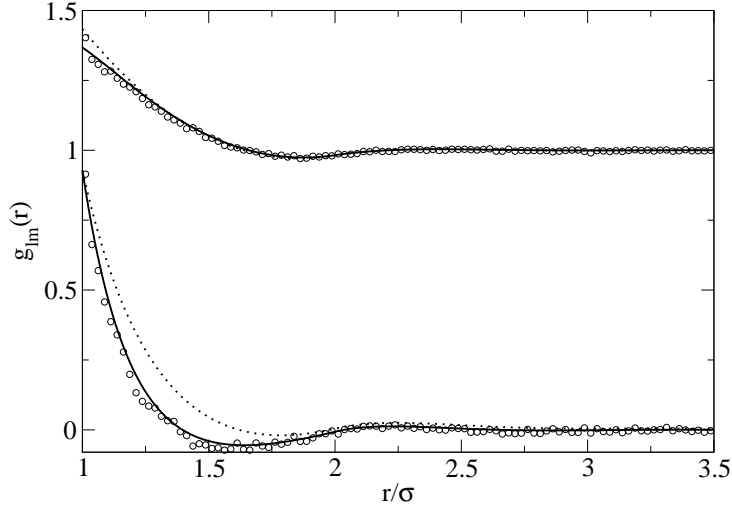}
\caption{\label{fig10} Same as Fig.\ref{fig9} for  $n^{*}=0.25$ and $a^{*}=1.0$.}
\end{figure*}

\begin{figure*}
\includegraphics[angle=0,scale=0.4]{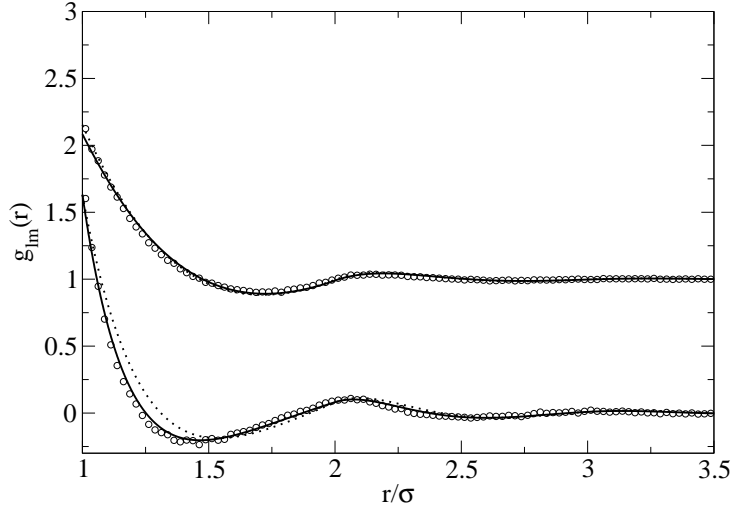}
\caption{\label{fig11} Same as Fig. \ref{fig9} for  $n^{*}=0.5$ and $a^{*}=0.6$.}
\end{figure*}

Figures \ref{fig9},\ref{fig10} and \ref{fig11} show $Re(g_{00}(r))$ and $Im(g_{22}(r))$ for $n^{\ast }=0.1$,$0.25$ and $0.5$ and $a^{\ast
}=1.0$, $1.0$ and $0.6$ respectively as determined from the simulations. While the spherically symmetric
component is little changed from equilibrium, the main nonequilibrium
component, $Im(g_{22}(r))$, is comparable in magnitude near the core to the equilibrium pdf but 
decays rapidly and is difficult
to measure beyond about two hard sphere diameters. Also shown in these
figures are the results of the GMSA model with both the Yukawa closure described
above and the simpler closure in which the full dcf is set equal to zero outside the core.
Both give a good description of the main features
of the structure including the location of the sign changes and the
amplitude and wavelength of the oscillations in $g_{22}$ with the Yukawa closure
being obviously superior in all cases. It is interesting that the 
largest discrepancy  occur for the lowest density. Based on the observed fluctuations in 
the identical determination of $g_{lm}(r)$ for this density in equilibrium, the statistical errors
for this system appear to account for no more that half the deviation seen in Fig.\ref{fig9}. One noticeable characteristic 
of the simulation data for $n^{*}=0.1$ is that $Im(g_{22})$ appears to be systematically below the GMSA model and indeed
systematically below zero away from the core. This suggests that the deviation might be a sign of the slow, algebraic decay predicted
by mode-coupling theory\cite{LutskoDufty-LongRangeShear} and which is not incorporated in the GMSA model.

\begin{figure*}
\includegraphics[angle=0,scale=0.4]{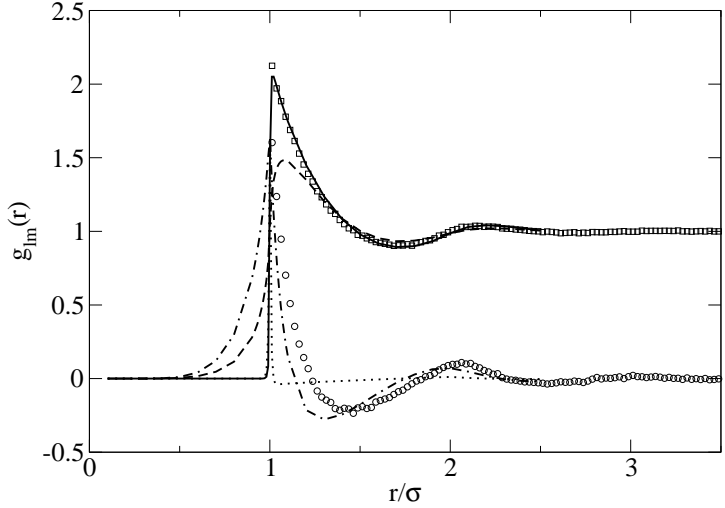}
\caption{\label{fig12}The functions $Re\left(g_{00}(r)\right)/\sqrt{4\pi}$ and 
$Im\left(g_{22}(r)\right)$ for $n^{*}=0.5$ and $a^{*}=0.6$ from simulation(squares and circles, respectively), the Hess model with the small er parameter (full and dotted lines, respectively) and with the larger parameter (dashed and dot-dash lines, respectively).}
\end{figure*}

\begin{figure*}
\includegraphics[angle=0,scale=0.4]{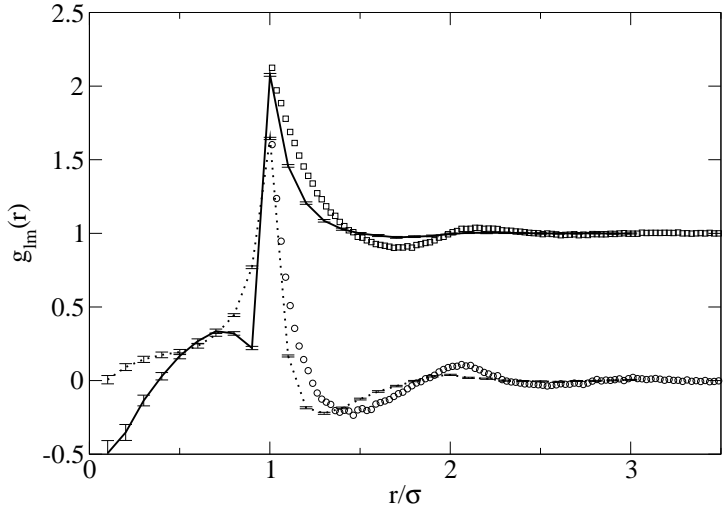}
\caption{\label{fig13}The functions $Re\left(g_{00}(r)\right)/\sqrt{4\pi}$ and 
$Im\left(g_{22}(r)\right)$ for $n^{*}=0.5$ and $a^{*}=0.6$ from simulation(squares and circles, respectively), and the Ronis model (full and dotted lines, respectively). The error bars are the standard errors reported by the VEGAS algorithm used to evaluate the model.}
\end{figure*}

Figures \ref{fig12} and \ref{fig13}  show the same simulation data as Fig.\ref{fig11} together with the numerical
evaluation of the models of Hess (eq.(\ref{hess1}))  and Ronis (eq.(\ref{ronis1})) performed by means of a Monte Carlo integration using 
the VEGAS algorithm\cite{Vegas,NR,GSL}. Because it can be formulated in real space, the Hess model requires one fewer integral than the Ronis model so that the evaluations are quicker and more accurate. In both cases, the parameters
were adjusted so as to reproduce the calculated value of $Im\left(M_{22}\right)$. Curiously, there are two values
of the free parameter in the Hess model that satisfy this constraint. For the smaller of the two values, 
$Re(g_{00}(r))$ is almost unchanged from equilibrium and $Im(g_{22}(r))$ is only nonzero in a very small region near the core.
Both components are non-zero in a small region inside the core. The larger value of the parameter gives rise to a
substantial deviation in $Re(g_{00}(r))$ which is therefore not in good agreement with the simulation data. 
In contrast, $Im(g_{22}(r))$ is in qualitative agreement with the data outside the core. In this case, both 
components take on substantial values inside the core. The Ronis model, shown in Fig. \ref{fig13}, is similar to the large-parameter 
version of the Hess model. The spherically symmetric component
is modeled somewhat better but $Im(g_{22}(r))$ is somewhat worse than with the Hess model. The behavior inside the core is also worse, with $Re(g_{00}(r))$ even taking on negative values.  
 
\bigskip

\section{Discussion}

The goal of the work presented here has been to describe the
density-density correlation function in a sheared hard-sphere fluid over a
range of densities, length scales and shear rates. It was shown that the
Enskog approximation for the velocity correlations provides a basis for
calculating the the density-density correlation function at contact and that
the results hold up well when compared to simulation, even for conditions of
moderate density and high shear rates. Indeed, deviations from the Enskog
predictions are primarily attributable to the high-shear rate phase
transition and, as noted previously\cite{Lutsko96}, appear to signal its
onset. For finite separations, the nonequilibrium GMSA was shown to provide
a framework within which the known information about the correlation
function at contact, coming from the Enskog approximation could be used to
provide an accurate description of the dominant effects at finite separations. While 
power-law decays arise naturally in the solution of the anisotropic OZ
equation, the inverse-cubic decay in $g_{22}(r)$ is in contradiction to the 
$1/r^{17/3}$ decay predicted by recent mode-coupling calculations\cite{LutskoDufty-LongRangeShear}.
Furthermore, those calculations indicate a $1/r^{11/3}$ decay in $g_{00}(r)$ for which
there is no analog in the OZ solution. Of course, such algebraic decays could find their 
origin in the auxiliary functions and, indeed, might be used as boundary conditions
in place of the Yukawa closure of the GMSA. On the other hand, the mode-coupling results
are the result of a number of approximations (linearized Navier-Stokes-Langevin model
solved perturbatively) and may not be giving the true asymptotic behavior.  
Altogether, the question of the actual form of these algebraic
decays must be considered to be unresolved at this point since the data
presented here is adequately fitted without them (except, possibly, for $n^{\ast} = 0.1$). More extensive simulations
which can provide better statistics for the decay at separations
significantly above two or three hard-sphere diameters would be required in
order to resolve this issue. What can be said here is that even if algebraic tails
exist, they must be significantly weaker than the dramatic nonequilibrium
contributions seen near the core.

Two other, closely related, theories for the pdf were also considered and
shown to capture some of the qualitative behavior of the pdf but both suffer
from the unphysical prediction of nonzero probabilities inside the core. 
Before dismissing this class of theory on this basis, it is interesting to
consider whether the main failing could be eliminated. In order to show that
this is indeed the case, consider the second equation of the BBGKY
hierarchy for hard spheres 
\begin{eqnarray}
&&\left[\frac{\partial }{\partial t}+\sum_{i=1,2}\left( 
\mathbf{p}_{i}^{\prime }\cdot \frac{\partial }{\partial \mathbf{q}_{i}}+%
\mathbf{q}_{i}\cdot \overleftrightarrow{a}\cdot \frac{\partial }{\partial 
\mathbf{q}_{i}}+\frac{\partial }{\partial \mathbf{p}_{i}^{\prime }}\cdot 
\mathbf{F}(x_{i})\right) +\overline{T}_{-}(12)\right]\rho
_{2}(x_{1},x_{2};t)  \label{bbgky2} \\
&=&-N\int d3\;\left( \overline{T}_{-}(13)+\overline{T}_{-}(23)\right) \rho
_{3}(x_{1},x_{2},x_{3};t) \notag
\end{eqnarray}%
First, observe that in the most general case the distribution must have the
form%
\begin{equation}
\rho _{2}(x_{1},x_{2};t)=\Theta \left( r_{12}-\sigma \right) \overline{\rho }%
_{2}(x_{1},x_{2};t)  \label{ansatz}
\end{equation}%
since the atoms cannot interpenetrate. Substituting this into Eq.(\ref%
{bbgky2}), one finds two terms proportional to $\delta \left( r_{12}-\sigma
\right) $ : the first coming from the action of the spatial gradients on
the step-function in eq.(\ref{ansatz}) and the second from the collisional
operator $\overline{T}_{-}(12)$. These must cancel so that their
coefficients must be equal and this just gives the relation between pre- and
post-collisional distribution functions used to derive Eq.(\ref{chaos1}).
The remaining regular part of the kinetic equation then reads%
\begin{eqnarray}
&&\Theta \left( q_{12}-\sigma \right) \left[ \frac{\partial }{\partial t}%
+\sum_{i=1,2}\left( \mathbf{p}%
_{i}^{\prime }\cdot \frac{\partial }{\partial \mathbf{q}_{i}}+\mathbf{q}%
_{i}\cdot \overleftrightarrow{a}\cdot \frac{\partial }{\partial \mathbf{q}%
_{i}}+\frac{\partial }{\partial \mathbf{p}_{i}^{\prime }}\cdot \mathbf{F}%
(x_{i})\right) \right]\overline{\rho }_{2}(x_{1},x_{2};t)\\
&=&-N\Theta \left( q_{12}-\sigma \right) \int d3\;\left( \overline{T}%
_{-}(13)+\overline{T}_{-}(23)\right) \overline{\rho }%
_{3}(x_{1},x_{2},x_{3};t)  \notag
\end{eqnarray}%
where an analogous decomposition of the 3-body distribution has been
introduced. Integrating over all momenta and discarding surface terms then
yields%
\begin{eqnarray}
&&\Theta \left( q_{12}-\sigma \right) \left[ \frac{\partial }{\partial t}%
+\mathbf{q}_{12}\cdot \overleftrightarrow{a}\cdot \frac{%
\partial }{\partial \mathbf{q}_{12}}\right] y(\mathbf{q}_{12};t)\\
&=&-\Theta \left( q_{12}-\sigma \right) n\int d\mathbf{p}%
_{1}d\mathbf{p}_{2}\int d3\;\left( \overline{T}_{-}(13) +\overline{T}%
_{-}(23)\right) \overline{\rho }_{3}(x_{1},x_{2},x_{3};t) \notag
\end{eqnarray}%
where $y(\mathbf{q}_{1},\mathbf{q}_{2};t)$ is the nonequilibrium cavity
function and is related to the pdf by $g(\mathbf{q}_{1},\mathbf{q}%
_{2};t)=\Theta \left( q_{12}-\sigma \right) y(\mathbf{q}_{1},\mathbf{q}%
_{2};t)$. This suggests that a more physical approximation would be to make
a relaxation approximation for the cavity function of the form 
\begin{equation}
\Theta \left( q_{12}-\sigma \right) \left[ \frac{\partial }{\partial t}y(%
\mathbf{q}_{12};t)+\mathbf{q}_{12}\cdot \overleftrightarrow{a}\cdot \frac{%
\partial }{\partial \mathbf{q}_{12}}y(\mathbf{q}_{12};t)-\int d\mathbf{r}\;A(%
\mathbf{q}_{12}-\mathbf{r})\left( y(\mathbf{r};t)-y_{0}(\mathbf{r})\right) %
\right] =0.
\end{equation}%
There is no reason at this point to keep the step-function in this equation
since any solution valid for all separations will be valid outside the
core. Then, this gives in Fourier space%
\begin{equation}
\frac{\partial }{\partial t}\widetilde{y}(\mathbf{k};t)-\mathbf{k}\cdot 
\overleftrightarrow{a}^{T}\cdot \frac{\partial }{\partial \mathbf{k}}%
\widetilde{y}(\mathbf{k};t)=\widetilde{A}(\mathbf{k})\left( \widetilde{y}(%
\mathbf{k};t)-\widetilde{y}_{0}(\mathbf{k})\right)
\end{equation}%
and for steady-state shear flow the solution is%
\begin{equation}
\widetilde{y}(\mathbf{k})=\widetilde{y}_{0}(k)+\int_{0}^{\infty }dt\left( \;%
\frac{ak_{x}k_{y}(-at)}{k}\widetilde{y}_{0}^{\prime }(k\left( -at\right)
)\right) \exp \left( -\int_{0}^{t}dt^{\prime }\;\widetilde{A}(\mathbf{k}%
\left( -at^{\prime }\right) )\right).
\end{equation}
An extension of the Hess model would be to take $\widetilde{A}\left(\mathbf{k}\right) = \sum_{lm}A_{lm}Y_{lm}(\widehat{k})$
for some set of constants $A_{lm}$ adjusted to give the correct moments at contact. A similar extension of the Ronis
model is also possible. An investigation of these models will be left to a future study.

In conclusion, it has been shown that the Enskog approximation for the pair correlations at contact, together with the GMSA model, 
provides a good description of the density autocorrelation function in a sheared fluid. The same techniques also give a good 
description of the pdf in granular fluids (modeled as inelastic hard spheres)\cite{LutskoHCS} giving evidence that the approach
is applicable to a variety of nonequilibrium systems. In both cases, simply knowing the value of the pdf at contact, 
from the Enskog approximation, and applying the standard formalism of liquid-state theory are enough to give a description 
of features of the system arising solely from the nonequilibrium state. Further work will include the
extension of this model to the description of static correlation functions involving temperature and velocity.
\bigskip 

\section{Acknowledgments}

The author acknowledges support from the Universit\'{e} Libre de
Bruxelles. 

\bigskip

\bigskip \appendix

\section{\protect\bigskip Solving the OZ model for USF}

\label{App:Exact}

The OZ equations written in terms of the auxiliary functions are 
\begin{equation}
\widetilde{h}_{lm}^{\prime }\left( k\right) =\widetilde{c}_{lm}^{\prime
}\left( k\right) +n\frac{1}{\sqrt{4\pi }}\sum_{\left| l^{\prime }-l^{\prime
\prime }\right| \leq l\leq l^{\prime }+l^{\prime \prime }}\sum_{m^{\prime
}=-l^{\prime }}^{l^{\prime }}A(l,l^{\prime },l^{\prime \prime },m,m^{\prime
})\widetilde{c}_{l^{\prime }m^{\prime }}^{\prime }\left( k\right) \widetilde{%
h}_{l^{\prime \prime }m-m^{\prime }}^{\prime }\left( k\right)  \label{a1}
\end{equation}%
and the boundary conditions, which follow directly from Eq.(\ref{aux}), are%
\begin{eqnarray}
\Theta (\sigma -r)h_{lm}^{\prime }(r) &=&-\sqrt{4\pi }\delta _{l0}\delta
_{m0}+(1-\delta _{l0})\sum_{n=0}^{\frac{l}{2}-1}B_{lm,n}r^{2n}  \label{a2} \\
\Theta (r-\sigma )c_{lm}^{\prime }(r) &=&v_{lm}^{\prime }(r)  \notag
\end{eqnarray}%
where the constant coefficients are functionals of the structure function%
\begin{equation}
\sum_{n=0}^{\frac{l}{2}-1}B_{lm,n}r^{2n}=-\frac{1}{r}\int_{\sigma }^{\infty
}dr^{\prime }\;\sum_{n=0}^{k-1}\left( 4n+3\right) P_{2n+1}\left( \frac{r}{%
r^{\prime }}\right) h_{lm}\left( r^{\prime };t\right) .  \label{a3}
\end{equation}%
Note that Eq.(\ref{a3}) is not a self-consistency condition: rather, it will
automatically be satisfied for any solution of eqs.(\ref{a1}) and (\ref{a2})
as can be verified using the relation%
\begin{equation}
h_{lm}(r;t)=h_{lm}^{\prime }\left( r;t\right) -\frac{1}{r^{2}}\sum_{n=0}^{%
\frac{l}{2}-1}\left( 4n+3\right) \int_{0}^{r}r^{\prime }dr^{\prime
}\;P_{2n+1}\left( \frac{r^{\prime }}{r}\right) h_{lm}^{\prime }\left(
r^{\prime };t\right) .
\end{equation}%
Instead, the significance of the constants is revealed by considering the
equivalent relation for the dcf%
\begin{equation}
c_{lm}(r;t)=c_{lm}^{\prime }\left( r;t\right) -\frac{1}{r^{2}}\sum_{n=0}^{%
\frac{l}{2}-1}\left( 4n+3\right) \int_{0}^{r}r^{\prime }dr^{\prime
}\;P_{2n+1}\left( \frac{r^{\prime }}{r}\right) c_{lm}^{\prime }\left(
r^{\prime };t\right)
\end{equation}%
which implies that%
\begin{eqnarray}
\Theta \left( r-1\right) c_{lm}(r;t) =\Theta \left( r-1\right)
v_{lm}^{\prime \prime }\left( r;t\right) - \\
\Theta \left( r-1\right)\frac{1}{r^{2}}\sum_{n=0}^{\frac{l%
}{2}-1}\left( 4n+3\right) \int_{0}^{1}r^{\prime }dr^{\prime
}\;P_{2n+1}\left( \frac{r^{\prime }}{r}\right) \left( c_{lm}^{\prime }\left(
r^{\prime };t\right) -v_{lm}^{\prime }\left( r^{\prime };t\right) \right) \notag%
\end{eqnarray}%
where%
\begin{equation}
v_{lm}^{\prime \prime }(r)=v_{lm}^{\prime }\left( r\right) -\frac{1}{r^{2}}%
\sum_{n=0}^{\frac{l}{2}-1}\left( 4n+3\right) \int_{0}^{r}r^{\prime
}dr^{\prime }\;P_{2n+1}\left( \frac{r^{\prime }}{r}\right) v_{lm}^{\prime
}\left( r^{\prime }\right) .
\end{equation}%
The reason for defining $v_{lm}^{\prime \prime }(r)$ is precisely due to the
non-uniqueness of this transformation, as discussed below Eq.(\ref%
{uniqueness}) of the text. The point is that if we assumed a closure of the
full dcf of the form $v_{lm}(r)=\phi _{lm}(r)+\sum_{n=1}^{\left[ \frac{l-1}{2%
}\right] }A_{lm,n}r^{-(2n+1)}$, then only $\phi _{lm}(r)$ would contribute
to $v_{lm}^{\prime }(r)$ and we would find that $v_{lm}^{\prime \prime
}(r)=\phi _{lm}(r)$. We would then complete the problem by adjusting the
constants $B_{lm,n}$ so that the boundary condition is satisfied, meaning 
\begin{equation*}
\sum_{n=1}^{\left[ \frac{l-1}{2}\right] }A_{lm,n}r^{-(2n+1)}=\frac{1}{r^{2}}%
\sum_{n=0}^{\frac{l}{2}-1}\left( 4n+3\right) \int_{0}^{r}r^{\prime
}dr^{\prime }\;P_{2n+1}\left( \frac{r^{\prime }}{r}\right) v_{lm}^{\prime
}\left( r^{\prime }\right) .
\end{equation*}%
However, in the present application, we are not concerned about the full dcf
and so the $B_{lm,n}$ are simply treated as free parameters.

If we retain only the $l=0,m=0$ and $l=2,m=\pm 2$, components, the model can
be reduced to the solution of a one dimensional OZ. The explicit form of
the OZ equations in this case are%
\begin{eqnarray}
\widetilde{h}_{00}^{\prime } &=&\widetilde{c}_{00}^{\prime }+n\sqrt{\frac{1}{%
4\pi }}\left[ \widetilde{c}_{00}^{\prime }\widetilde{h}^{\prime
}{}_{00}+\left( \widetilde{c}_{22}^{\prime }\widetilde{h}_{2-2}^{\prime }+%
\widetilde{c}_{2-2}^{\prime }\widetilde{h}_{22}^{\prime }\right) \right] \\
\widetilde{h}_{22}^{\prime } &=&\widetilde{c}_{22}^{\prime }+n\sqrt{\frac{1}{%
4\pi }}\left( \widetilde{c}_{00}^{\prime }\widetilde{h}_{22}^{\prime }+%
\widetilde{c}_{22}^{\prime }\widetilde{h}_{00}^{\prime }\right)  \notag \\
\widetilde{h}^{\prime }{}_{2-2} &=&\widetilde{c}_{2-2}^{\prime }+n\sqrt{%
\frac{1}{4\pi }}\left( \widetilde{c}_{00}^{\prime }\widetilde{h}%
_{2-2}^{\prime }+\widetilde{c}_{2-2}^{\prime }\widetilde{h}_{00}^{\prime
}\right)  \notag
\end{eqnarray}%
with the boundary conditions%
\begin{eqnarray}
\Theta (\sigma -r)h_{00}^{\prime }(r) &=&-\sqrt{4\pi } \\
\Theta (\sigma -r)h_{2\pm 2}^{\prime }(r) &=&-B_{2\pm 2,0}  \notag \\
\Theta (r-\sigma )c_{00}^{\prime }(r) &=&v_{00}^{^{\prime }(0)}(r)  \notag \\
\Theta (r-\sigma )c_{2\pm 2}^{\prime }(r) &=&v_{22}^{^{\prime }(0)}(r) 
\notag \\
\lim_{\epsilon \rightarrow 0}h_{00}^{\prime }(\sigma +\epsilon ) &=&M_{00} 
\notag \\
\lim_{\epsilon \rightarrow 0}h_{2\pm 2}^{\prime }(\sigma +\epsilon )
&=&M_{2\pm 2}+B_{2\pm 2,0}  \notag
\end{eqnarray}%
First, note that $h_{l-m}=(-1)^{m}h_{lm}^{\ast }$, etc. which follows from
the equivalent property of the spherical harmonics. Then, it is useful to
separate the equations into their real and imaginary parts to get%
\begin{eqnarray}
\widetilde{h}_{00}^{\prime } &=&\widetilde{c}_{00}^{\prime }+n\sqrt{\frac{1}{%
4\pi }}\left[ \widetilde{c}_{00}^{\prime }\widetilde{h}^{\prime
}{}_{00}+2\left( \widetilde{c}_{22,r}^{\prime }\widetilde{h}_{22,r}^{\prime
}+\widetilde{c}_{22,i}^{\prime }\widetilde{h}_{22,i}^{\prime }\right) \right]
\\
\widetilde{h}_{22,r}^{\prime } &=&\widetilde{c}_{22,r}^{\prime }+n\sqrt{%
\frac{1}{4\pi }}\left( \widetilde{c}_{00}^{\prime }\widetilde{h}%
_{22,r}^{\prime }+\widetilde{c}_{22,r}^{\prime }\widetilde{h}_{00}^{\prime
}\right)  \notag \\
\widetilde{h}^{\prime }{}_{22,i} &=&\widetilde{c}_{22,i}^{\prime }+n\sqrt{%
\frac{1}{4\pi }}\left( \widetilde{c}_{00}^{\prime }\widetilde{h}%
_{22,i}^{\prime }+\widetilde{c}_{22,i}^{\prime }\widetilde{h}_{00}^{\prime
}\right)  \notag
\end{eqnarray}%
where $\widetilde{h}_{22,r}^{\prime } \equiv Re\left(\widetilde{h}_{22}^{\prime }\right)$, etc.
Second, because of the linearity in the $22$-components of the last two
equations and the boundary conditions on the core-values of the
components of the structure function are constants, these equations are solved by taking%
\begin{eqnarray}
\widetilde{h}_{22,r}^{\prime } &=&x\widetilde{h}^{\prime }{}_{22,i} \\
x &=&M_{22,r}/M_{22,i}  \notag
\end{eqnarray}%
giving%
\begin{eqnarray}
\widetilde{h}_{00}^{\prime } &=&\widetilde{c}_{00}^{\prime }+n\sqrt{\frac{1}{%
4\pi }}\left[ \widetilde{c}_{00}^{\prime }\widetilde{h}^{\prime
}{}_{00}+2\left( 1+x^{2}\right) \widetilde{c}_{22,i}^{\prime }\widetilde{h}%
_{22,i}^{\prime }\right] \\
\widetilde{h}_{22,r}^{\prime } &=&\widetilde{c}_{22,r}^{\prime }+n\sqrt{%
\frac{1}{4\pi }}\left( \widetilde{c}_{00}^{\prime }\widetilde{h}%
_{22,r}^{\prime }+\widetilde{c}_{22,r}^{\prime }\widetilde{h}_{00}^{\prime
}\right)  \notag
\end{eqnarray}%
provided $Re\left(v^{0}_{22}\right)=xIm\left(v^{0}_{22}\right)$ which we are free
to impose. Now, define $h(r;u)=h_{00}^{\prime }(r)+uh_{22,i}^{\prime }(r)$, which
satisfies%
\begin{eqnarray}
\widetilde{h}(k,u) &=&\widetilde{c}(k,u)+n\sqrt{\frac{1}{4\pi }}\left[ 
\widetilde{c}_{00}^{\prime }\widetilde{h}^{\prime }{}_{00}+u\left( 
\widetilde{c}_{00}^{\prime }\widetilde{h}_{22,r}^{\prime }+\widetilde{c}%
_{22,r}^{\prime }\widetilde{h}_{00}^{\prime }\right) +2\left( 1+x^{2}\right) 
\widetilde{c}_{22,i}^{\prime }\widetilde{h}_{22,i}^{\prime }\right] \\
&=&\widetilde{c}(k,u)+n\sqrt{\frac{1}{4\pi }}\widetilde{c}(k,u)\widetilde{h}%
(k,u)  \notag
\end{eqnarray}%
where the last line follows if and only if $u=\pm \sqrt{2\left(
1+x^{2}\right) }$. The boundary conditions are then%
\begin{eqnarray}
\Theta (\sigma -r)h(r;u) &=&-\sqrt{4\pi }-u\func{Im}\left( B_{22,1}\right) \\
\Theta (r-\sigma )c(r,u) &=&v_{00}^{^{\prime }(0)}(r)+uv_{22}^{^{\prime
}(0)}(r)  \notag
\end{eqnarray}%
Finally, introduce scaled quantities defined by $h(r;u)=\left[ \sqrt{4\pi }\pm\left| u \right|%
\func{Im}\left( B_{22,1}\right) \right] H_{\pm}(r)$ and $c(r;u)=\left[ \sqrt{%
4\pi }+\left| u \right| \func{Im}\left( B_{22,1}\right) \right] C_{\pm}(r)$ which give%
\begin{eqnarray}
\widetilde{H}_{\pm }(k) &=&\widetilde{C}_{\pm }(k)+n_{\pm }\widetilde{C}%
_{\pm }(k)\widetilde{H}_{\pm }(k) \\
\Theta (\sigma -r)H_{\pm }(r) &=&-1  \notag \\
\Theta (r-\sigma )C_{\pm }(r) &=&\left[ \sqrt{4\pi }\pm \left| u\right| 
\func{Im}\left( B_{22,1}\right) \right] ^{-1}v_{00}^{^{\prime }(0)}(r)\pm
\left| u\right| \left[ \sqrt{4\pi }\pm \left| u\right| \func{Im}\left(
B_{22,1}\right) \right] ^{-1}v_{22}^{^{\prime }(0)}(r)  \notag
\end{eqnarray}%
which can be recognized as the OZ equation for a (possibly negative)\
density $n_{\pm }=n\left[ 1\pm \left| u\right| \sqrt{\frac{1}{4\pi }}\func{Im%
}\left( B_{22,1}\right) \right] $ and some particular closure condition (so
that this resembles the usual GMSA). For the simplest case, in which one
takes $v_{00}^{^{\prime }(0)}(r)=0$, we have $H_{\pm }(r)=h_{py}(r;n_{\pm })$
and the solution is trivial. If the tail functions,$v_{00}^{^{\prime
}(0)}(r) $ and $v_{22}^{^{\prime }(0)}(r)$ are Yukawas, then we can use of
the solution of Hoye and Blum for a closure consisting of a sum of Yukawas\cite%
{Hoye77}. For completeness, we collect together the various transformations
to see that%
\begin{eqnarray}
h_{00}^{\prime }(r) &=&\frac{\sqrt{4\pi }}{2n}\left(
n_{+}H_{+}(r)+n_{-}H_{-}(r)\right)
\\
h_{22,i}^{\prime }(r) &=&\frac{\sqrt{4\pi }}{2\left| a\right| n}\left(
n_{+}H_{+}(r)-n_{-}H_{-}(r)\right)
\notag \\
h_{22,r}^{\prime }(r) &=&xh_{22,i}^{\prime }(r)  \notag
\end{eqnarray}%
. The value of $\func{Im}\left( B_{22,1}\right) $ is, of course, fixed by
requiring that 
\begin{equation}
\lim_{\epsilon \rightarrow 0+}h_{22,i}^{\prime }(\sigma +\epsilon )=-\func{Im%
}\left( B_{22,1}\right) +\func{Im}\left( M_{22}\right)
\end{equation}%
while the full pdf is%
\begin{eqnarray}
h(\mathbf{r}) &=&h_{00}^{\prime }(r)Y_{00}(r)+2h_{22,r}^{\prime \prime }(r)%
\func{Re}Y_{22}(\widehat{r})-2h_{22,i}^{\prime \prime }(r)\func{Im}Y_{22}(%
\widehat{r}) \\
&=&\frac{1}{\sqrt{4\pi }}h_{00}^{\prime \prime }(r)+\sqrt{\frac{5}{8\pi }}%
h_{22,r}^{\prime \prime }(r)\left( \widehat{r}_{x}^{2}-\widehat{r}%
_{y}^{2}\right) -\sqrt{\frac{5}{2\pi }}h_{22,i}^{\prime \prime }(r)\widehat{r%
}_{x}\widehat{r}_{y}  \notag
\end{eqnarray}%
where%
\begin{eqnarray}
h_{22}^{\prime \prime }(r) &=&\int_{0}^{\infty }r^{\prime 2}dr^{\prime
}\;U_{2}(r^{\prime },r)h_{22}^{\prime }\left( r^{\prime };t\right) \\
&=&\Theta \left( r-1\right) \left[ h_{22}^{\prime }\left( r^{\prime
};t\right) +\frac{\func{Im}\left( B_{22,1}\right) }{r^{3}}-\frac{3}{r^{3}}%
\int_{1}^{r}r^{\prime 2}dr^{\prime }\;h_{22}^{\prime }\left( r^{\prime
};t\right) \right]  \notag
\end{eqnarray}

\bigskip

\bibliography{physics}

\end{document}